\documentclass[reprint, prx]{revtex4-2}
\usepackage{amssymb}
\usepackage{amsmath}
\usepackage{braket}
\usepackage{color}
\usepackage{MnSymbol}
\usepackage{ulem}
\usepackage{color}
\usepackage{comment}
\usepackage{graphicx}
\usepackage{hyperref}
\usepackage[only,llbracket,rrbracket,llparenthesis,rrparenthesis]{stmaryrd}

\usepackage{leftindex}

\DeclareMathOperator{\spec}{spec}

\newcommand{\be}{\begin{equation}}
\newcommand{\ee}{\end{equation}}
\newcommand{\ba}{\begin{aligned}}
\newcommand{\ea}{\end{aligned}}
\newcommand{\bs}[1]{#1}

\newcommand{\HS}{\chi}

\newcommand{\GS}{\textsc{gs}}

\newcommand{\tr}{\mathrm{tr}}

\begin{document}

\title{
Disorder-Order Interface Propagating over the Ferromagnetic Ground State\\
in the Transverse Field Ising Chain
}
\author{Vanja Mari\'c}
\author{Florent Ferro}
\author{Maurizio Fagotti}
\affiliation{Universit\'e Paris-Saclay, CNRS, LPTMS, 91405, Orsay, France}

\date{\today}

\begin{abstract}
We consider time evolution of order parameters and entanglement asymmetries in the ferromagnetic phase of the transverse-field Ising chain. One side of the system is prepared in a ferromagnetic ground state and the other side either in equilibrium at higher temperature or out of equilibrium. We focus on the disorder-order interface in which the order parameter attains a nonzero value, different from the ground state one. 
In that region, correlations follow a universal behaviour. We analytically compute the asymptotic scaling functions of the one- and two-point equal time correlations of the order parameter and provide numerical evidence that also the non-equal time correlations are universal. 
We analyze the R\'enyi entanglement asymmetries of subsystems and obtain a prediction that is expected to hold also in the von Neumann limit.
Finally, we show that the Wigner-Yanase skew information of the order paramerter in subsystems within the interfacial region scales as their length squared. We propose a semiclassical approximation that is particularly effective close to the edge of the lightcone.  
\end{abstract}
\maketitle

\tableofcontents

\section{Introduction}

Isolated quantum many-body systems cannot relax as a whole, but they can relax locally in the thermodynamic limit~\cite{Polkovnikov2011Colloquium,Gogolin2016Equilibration}. A compelling argument for this is that the system behaves as its own bath: the lack of information from distant parts of the system can be effectively treated as though the accessible region were coupled to an external reservoir. In integrable one dimensional systems, such intuition holds true in a rigorous sense. Consider, for instance, the time evolution governed by a translationally invariant Hamiltonian in a system with a thermodynamic Bethe Ansatz description. When the system is bipartitioned, with each half independently prepared either in thermal equilibrium or in the ground state of two potentially different {Hamiltonians}, its dynamics can be described by generalized hydrodynamics (GHD)~\cite{Castro-Alvaredo2016Emergent,Bertini2016Transport,Borsi2020,Alba2021Generalized,DeNardis2022Correlation,Bulchandani2021Superdiffusion,Borsi2021Current,Essler2023review}. The latter consists of integro-differential equations for a type of Wigner functions, which in the specific research field are called ``root densities''.
The boundary conditions for the GHD equation(s) are not in one-to-one correspondence with the state at the initial time, which, indeed, is not fully characterised by the hydrodynamic quantities.  Instead, these boundary conditions reflect the thermodynamic stationary states that describe local observables at late times outside the lightcone spreading from the junction. In other words, the regions outside the lightcone act as an effective reservoir for those inside it.

The aptitude of generalised hydrodynamics to remain unaffected by the details of the initial state, even when those details are complex yet ultimately irrelevant at late times, makes it a very powerful theory. However, this same feature also makes it subject to subtleties that may compromise the completeness of the description,  and they are not always immediately apparent. One such issue was recognized shortly after the first GHD proposals: in the easy-axis XXZ spin-$\frac{1}{2}$ chain---an interacting integrable system---the root densities do not determine the sign of the expectation values of operators that are odd under spin-flip~\cite{Ilievski2016}. Ref.~\cite{Piroli2017Transport} proposed complementing the root-density characterization with a binary variable to account for the sign of those expectation values. This refined GHD description predicts sharp changes in odd observables over a length scale negligible compared to the typical scale over which the expectation values of even local observables vary. Similar, though distinct, issues also arise in noninteracting models. For example, consider the protocols studied by Eisler and collaborators in Refs~\cite{Eisler2016Universal,eisler2020Front,Eisler2018Hydrodynamical}. Two ground states of a system in an ordered phase are joined, with the underlying symmetry broken in different ways. At late times, the expectation values of local observables become nontrivial functions of the distance from the junction per unit time. What does GHD predict? Nothing. This time the issue is that the root densities are blind to symmetry breaking, hence GHD, as a stand-alone theory, cannot distinguish  macroscopically different ground states from one another, and, in turn,  is clearly unable to predict the late time behaviour of local observables when  two different ground states are joined.

While such issues could  be the result of some deficiency in the original GHD description, we argue that they could also signal unusual physical properties of the state.
We have indeed recently confirmed our suspicions of anomalous behaviours in the presence of spontaneous symmetry breaking~\cite{Maric2024Macroscopic,Ferro2025Kicking}. 
Specifically, in Ref.~\cite{Maric2024Macroscopic} we considered the bipartitioning protocol in which a part of the state is prepared in a symmetry breaking ground state and the other in equilibrium at higher temperature. We have shown that both classical and quantum correlations do not cluster inside the interfacial region in which the order parameter varies from zero to the ground state value, and we announced that they are described by universal scaling functions. To understand the anomaly, let 
$A_{x(t)}$ be a subsystem centered at  $x(t)$, whose reduced density matrix can be approximated by that of a homogeneous stationary state 
$\rho^{\mathrm{st}}_{x(t),t}$ under a certain criterion of similarity.  
Ref.~\cite{Maric2024Macroscopic} suggests that, if $A_{x(t)}$ is in the interfacial region, it is impossible to choose $\rho^{\mathrm{st}}_{x(t),t}$ in a way that correctly predicts {both} the behavior of local observables around $x(t)$ { and the connected correlations that remain significantly different from zero even with observables outside 
$A_{x(t)}$.}

We prove here the results announced in Ref.~\cite{Maric2024Macroscopic} in the specific case of the transverse-field Ising model and extend the analysis to the entanglement asymmetry. 
We also prove that the late time behaviour is not affected by localized perturbations at the initial time, which is not a priori obvious since the effect of localized perturbations over a symmetry-breaking ground state does not generally fade away~\cite{eisler2020Front}. 
Finally, we investigate quantum correlations using the Wigner-Yanase skew information and propose a semiclassical theory that provides a good approximation of their scaling functions in a particular regime. 
While our worked example is the two-temperature scenario in the transverse-field Ising chain considered in Ref.~\cite{Maric2024Macroscopic}, we show that the only relevant aspect of the protocol is that half of the initial state is prepared in a symmetry-breaking ground state, provided that the other part exhibits or develops an extensive entropy.  

\subsection{Protocol}

The transverse-field Ising chain is described by the Hamiltonian
\begin{equation}\label{eq:HIsing}
    \bs H^{(h)} =-\sum_{\ell}( \bs\sigma^x_{\ell} \bs\sigma^x_{\ell+1}+h \bs\sigma^z_\ell)\, , 
\end{equation}
where $\bs \sigma^\alpha_\ell$ for $\alpha=x,y,z$ act as Pauli matrices on site $\ell$ and as the identity elsewhere; $h$ parametrises the effect of an external magnetic field in the transverse direction.
For $|h|<1$ the model exhibits a zero-temperature ferromagnetic phase in which the spin-flip symmetry associated with the transformation $\prod_j\bs\sigma_j^z$ is spontaneously broken. Namely, for $|h|<1$ there are two stable ground states $\ket{\textsc{gs}_\pm}$ with spontaneous magnetization~\cite{Sachdev2011book}
\begin{equation}
    \braket{\textsc{gs}_\pm|\bs \sigma^x_\ell |\textsc{gs}_\pm}=\pm m^x_{\GS} \; ,
\end{equation}
where
\begin{equation}
    m^x_{\GS}= (1-h^2)^{1/8} \; .
\end{equation}
For $|h|>1$, instead, the model exhibits a paramagnetic phase in which spins tend to align in the transverse direction.

In most of the paper, we study the bipartitioning protocol in which the system is prepared in an equilibrium state for the split Ising Hamiltonian $\bs H_0^{(h,h)}$, with $0<h<1$, in which the coupling between site $0$ and $1$ is turned off, where
\begin{equation}\label{eq:H0}
    \bs H_0^{(h_l,h_r)} =-\sum_{\ell\leq 0}[ \bs\sigma^x_{\ell-1} \bs\sigma^x_{\ell}+h_l \bs\sigma^z_\ell]-\sum_{\ell> 0}[ \bs\sigma^x_{\ell} \bs\sigma^x_{\ell+1}+h_r \bs\sigma^z_\ell]\; .
\end{equation} 
We denote by $\beta$ the inverse temperature of the left part; the right part is prepared at zero temperature.
The spin-flip symmetry is broken on the right hand side, where  the longitudinal magnetization (far enough from the boundary) is equal to $m_{\textsc{gs}}^x$.
We then consider the local quench consisting in switching on the coupling that  was originally off, that is to say, the state time evolves under Hamiltonian~\eqref{eq:HIsing}: $\bs H_0^{(h,h)}\rightarrow\bs H^{(h)}$. 

Alternatively, we consider the global quench in which the left part of the state is prepared in an equilibrium state (e.g., the ground state) of a different Hamiltonian, for example $\bs H_0^{(h_l,h)}\rightarrow\bs H^{(h)}$.

\begin{figure}[t]
    \centering
    \includegraphics[width=1.\linewidth]{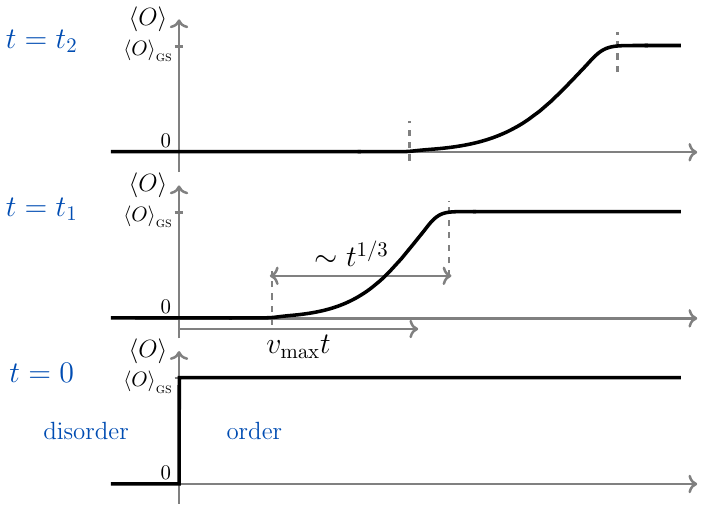}
    \caption{{Schematic representation of the profile of an order parameter $\braket{\bs O}$ at the initial time $t=0$ and two subsequent times $t=t_1,t_2$ after the quench; $\braket{\bs O}_{\textsc{GS}}$ denotes the expectation value of the operator $\bs O$ in the symmetry-breaking ground state. }The interfacial region moves at the maximal velocity of the quasiparticle excitations and spreads as $t^{1/3}$.}
    \label{fig:schematic}
\end{figure}

\subsection{Review of established results}
Since the two-temperature bipartitioning protocol in the transverse-field Ising chain has been widely investigated, we briefly review some of the results obtained in the past.  

The local relaxation to a nonequilibrium steady state was firstly addressed in the framework of $C^\star$ algebra~\cite{Aschbacher2003}, where it was shown, in particular, that, for strictly nonzero temperatures, the stationary state (NESS) capturing the expectation value of local observables in the limit of infinite time does not depend on the details of the initial state around the junction of the two parts prepared in thermal equilibrium. The same problem was independently studied in Ref.~\cite{Platini2005Scaling} in large but finite chains, and the authors pointed out an emerging scaling limit for the expectation values of the transverse spin, which turned out to become a function of $\ell/t$, with $\ell$ the distance from the junction. They also noted that at the edge of the lightcone the thermal expectation value is approached as $t^{-1/3}$ over a scale of order $t^{1/3}$, but they didn't realise the connection with the Airy kernel introduced by Tracy and Widom a decade before~\cite{Tracy1993Level}. We also mention Refs~\cite{Hunyadi2004,Eisler2013} that obtained analogous results in other noninteracting systems. To the best of our knowledge, the connection with the Airy kernel in the quantum Ising model was highlighted much later~\cite{Perfetto2017Ballistic,Fagotti2017Higher}. Some universal aspects of the NESS at the critical point ($h=1$ in Eq.~\eqref{eq:HIsing}) have been pointed out in \cite{Bernard2012Energy} in the framework of conformal field theory. In particular, the authors computed the energy full counting statistics, showing that, besides the temperatures of the initial state, it depends only on the central charge of the conformal field theory. In a larger collaboration focussed on the quantum Ising chain~\cite{DeLuca2013Nonequilibrium}, they confirmed the CFT results and detailed the heat transport and the energy full counting statistics in the paramagnetic phase ($h>1$). A similar, more extensive, investigation was carried out in Refs \cite{Perfetto2017Ballistic,Perfetto2020Dynamics}. Arguably, \cite{Kormos2017} is the most complete reference for the expectation value of local operators and equal-time two-point spin correlation functions in the quantum Ising model after the junction of two macroscopically different semi-infinite chains, such as the two-temperature scenario under consideration.

An unsuccessful attempt to reformulate the problem in the framework of phase-space quantum mechanics was reported in \cite{Fagotti2017Higher}. That task was eventually accomplished in \cite{Fagotti2020}, which, on the one hand,  reinterpreted and enhanced the generalised hydrodynamic equation as the Schr\"odinger equation in an invariant manifold, named ``locally quasistationary states'' after Ref.~\cite{Bertini2016Determination}, and, on the other hand, worked out the phase-space formulation of dynamics in noninteracting spin chains starting from Gaussian initial states. The theory  was then generalised in Ref.~\cite{Bocini2023Connected} to include non-Gaussian states with spin-flip symmetry. There, Bocini pointed out also some limitations of generalised hydrodynamics on capturing the asymptotic behaviour of connected two-point functions. 

The exceptional phenomenology in the presence of spontaneous symmetry breaking had been pointed out in \cite{Zauner2015,Eisler2016Universal}. As far as we can tell, however, such works went almost unnoticed, despite being in apparent conflict with the flourishing theory of generalised hydrodynamics, which, in a noninteracting model like the Ising one, was supposed to provide a complete description of the dynamics. The apparent conflict with GHD is probably more evident when considering the situation later discussed in Ref.~\cite{eisler2020Front}. There it was shown that, close to the ground state, the expectation values of local observables at the Euler scale, homeland of generalised hydrodynamics, turn out to be highly sensitive to microscopic details of the initial state that become invisible at mesoscopic scales, making it even unclear how they could be fed into a hydrodynamic equation. 
The issue of the dependence of the late-time behaviour on the details of the junction of two ground states with spontaneous symmetry breaking has been also recently addressed in  Ref.~\cite{Delfino2022Space} in the continuum scaling limit in which the model can be described by a quantum field theory. 

We conclude with a list of features already observed:
\begin{itemize}
\item In the limit of large time, the expectation values of spin-flip invariant local observables approach functions of $\frac{\ell}{t}$, where $\ell$ is the position of the observable with respect to the junction. 
\item The asymptotic behaviour of spin-flip invariant local observables in the limit of large time does not depend on how the initial thermal states have been joined.  
\item At the edges of the lightcone representing the region reached by the information about the junction of the initial thermal states, spin-flip invariant local observables approach their thermal expectation values as $t^{-\frac{1}{3}}$ over a region of order $t^{\frac{1}{3}}$. Their behaviour has an underlying connection with the Airy kernel, which can be understood within the theory of generalised hydrodynamics as a third-order effect in the formal expansion in spatial derivatives~\cite{Alba2021Generalized}. 
\end{itemize}

\section{Results}

The first aspect we point out is that 
\begin{description}
\item[Order is destroyed at the maximal velocity] 
This has a very simple interpretation based on the fact that, in one dimension, order is not expected in a stationary state with extensive entropy.
\end{description}
From  Ref.~\cite{Bertini2018Entanglement} it follows that the entropy of spin blocks is extensive in the limit of infinite time along any ray $\zeta= \ell/t$  with $\zeta$ strictly smaller than the maximal velocity of the excitations. Thus, non-symmetric observables can have a nonzero expectation value only starting from around the right edge of the lightcone. 

In the initial state that we  consider, the transition from ferromagnetic order to disorder is  abrupt, but time evolution smooths out the region at the interface. In particular we find

\begin{description}
\item[The width of the interfacial region scales as $t^{1/3}$] 
It corresponds to the region in which the behaviour of spin-flip invariant observables is characterised by the Airy kernel.
\end{description}
The interfacial region is schematically represented in Fig.~\ref{fig:schematic}. We remark that, while symmetric observables have expectation values close to their values outside the lightcone (the discrepancy is $O(t^{-1/3})$), the expectation value of non-symmetric observables, such as the local longitudinal spin, ranges from $0$ to their value in the ground state. 

\begin{figure}[htbp]
    \includegraphics[width=\linewidth]{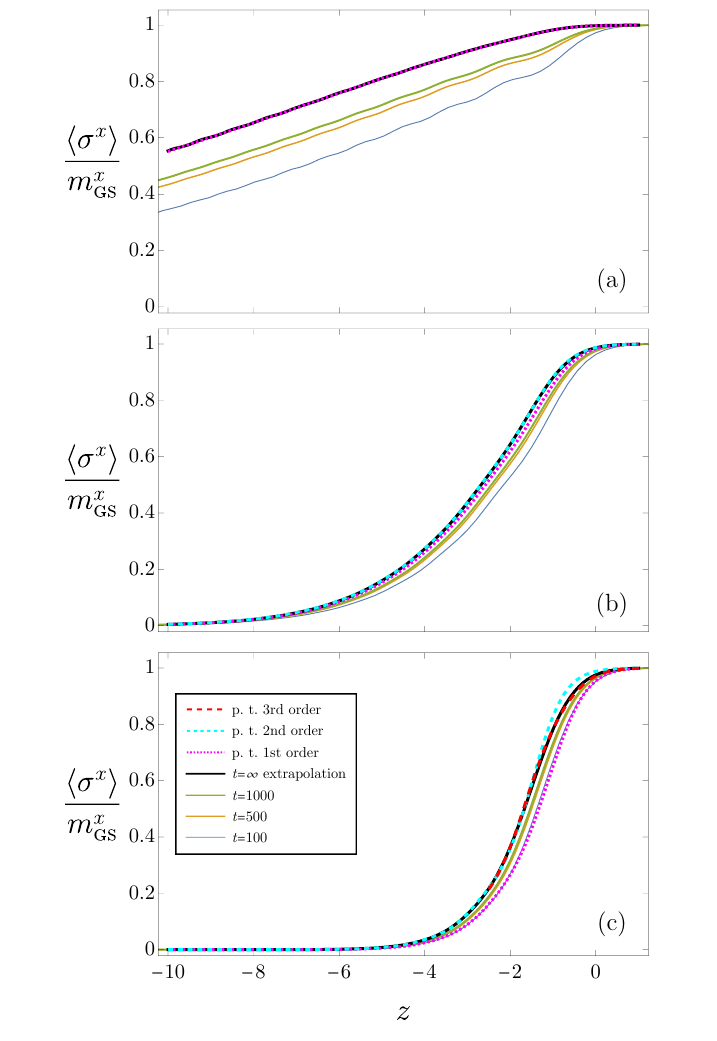}
    \caption{Magnetization $\braket{\bs \sigma^x_j}$ near the right edge of the lightcone, as a function of $z=\tfrac{j-v_{\rm max}t}{|t v''(\bar p)|^{\frac{1}{3}}}$,  at different times $t$ after the quench, including the extrapolation to $t\to\infty$ {(see Section~\ref{sec:numerical} for details)}. The magnetic field is set to $h=0.5$ and the inverse temperature of the left thermal reservoir is a) $\beta=2$, b) $\beta=0.75$, c) $\beta=0.25$. The data are compared with the predictions at the lowest orders of the perturbation theory.}
    \label{fig:mx}
\end{figure}

\begin{figure}[t]
    \centering
  \includegraphics[width=1\linewidth]{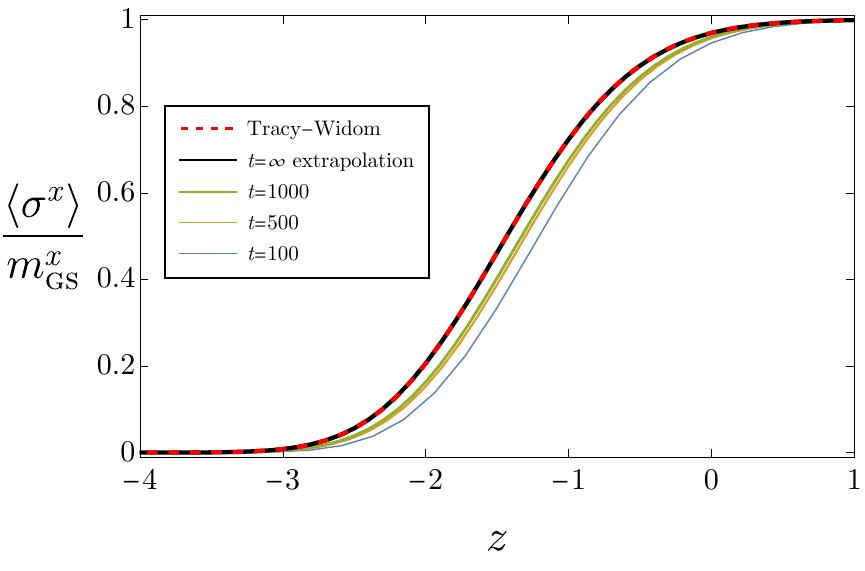}
    \caption{The same as in Fig.~\ref{fig:mx} for $\beta=0$, with the magnetization rescaled by its ground state value. The prediction (dashed red line) is the Tracy-Widom distribution $F_2(2^{1/3}z)$.}
    \label{fig:magnetization_infinite_temperature}
\end{figure}

The local longitudinal magnetization $\braket{\sigma^x}$ is the conventional order parameter in the Ising model{, which is particularly simple since it can be mapped to a chain of free fermions}. 
In a Gaussian state {(a state in which all correlations are determined by the 2-point fermion correlations via the Wick's theorem)}, it can be reduced to a determinant (in fact, a Pfaffian) of a half-infinite matrix constructed with the elements of the correlation matrix.  In the setting we consider the size of the matrix can be effectively reduced because the state is homogeneous outside the lightcone, where the order parameter attains a nonzero value. Close to the right edge of the lightcone such an effective size is then of order $t^{1/3}$.
This technical insight suggests that $\braket{\sigma^x}$ is highly sensitive only to perturbations to the initial state that alter the elements of the correlation matrix around the right edge of the lightcone by $\sim t^{-\frac{1}{3}}$ contributions. As also discussed in Ref.~\cite{Alba2021Generalized}, localised perturbations that keep the initial state Gaussian affect the edge of the lightcone with $\sim t^{-2/3}$ corrections. {This suggests that localised perturbations do not affect the edge of the lightcone. It is worth contrasting this finding with that of Ref.~\cite{Eisler2016Universal} (cf. also  Ref.~\cite{Ferro2025Kicking}), which studied the same model but with the left and right part prepared in  symmetry breaking ground states. They established a strong sensitivity to localized perturbations of the magnetization profile, which is found to vary over ballistic scales. In our protocol, instead, the magnetization changes abruptly at the edge of the lightcone, and the important sensitivity to local perturbations pointed out in Ref.~\cite{Eisler2016Universal} becomes just a $~t^{-1/3}$ correction---see Section~\ref{Sec:order parameter correlations} for a proof of the stability of asymptotic result under localized perturbations.}
\begin{description}
\item[The asymptotics of $\braket{\sigma^x}$ are universal] they do not depend on microscopic details of the initial state.
\end{description}
The order parameter's profile in the interfacial region depends on the left reservoir only through the density of excitations with the maximal velocity, which we parametrize as $\frac{1-e^{-\eta}}{2}$. In the thermal case that we consider
\begin{equation}\label{eta explicit}
    \eta=-\log \left[\tanh\left(\tfrac{\beta\varepsilon(\bar p)}{2}\right)\right]=-\log \left[\tanh\left(\beta\sqrt{1-h^2}\right)\right] \ ,
\end{equation}
where $\varepsilon(p)=2\sqrt{1+h^2-2h\cos p}$ is the dispersion relation and $\bar p$ is the momentum of the excitation with maximal velocity.
We find
\begin{equation}\label{eq:sx}
\braket{\bs \sigma_j^x}_t=m^x_{\GS} \mathcal M_\eta \Bigl(\tfrac{j-v_{\rm max}t}{|t v''(\bar p)|^{\frac{1}{3}}}\Bigr)+O(t^{-\frac{1}{3}})\, ,
\end{equation}
where $v(p)=\varepsilon'(p)$ is the velocity of the quasiparticle excitations with momentum $p$, {$v_{\rm max}=\max_p v(p)=v(\bar p)$}, and $\mathcal M_\eta$ is a universal scaling function that depends on the details of the left reservoir only through $\eta$. Within the framework of Ref.~\cite{Fagotti2024asymptotic}, $\mathcal M_\eta$ could be  expressed in terms of the factors of the Wiener-Hopf star factorizations of the symbol
\begin{equation}\label{eq:symboledge}
a(z,q)=1-(1-e^{-\eta})\HS(2z+q^2)\,  \qquad z,q\in\mathbb R\, .
\end{equation}
Here $\HS$ is a primitive of the Airy function, $
\HS(x)=\pi[\mathrm{Ai}(x)\mathrm{Gi}'(x)-\mathrm{Gi}(x)\mathrm{Ai}'(x)]
$, and $\mathrm{Gi}(x)$ is one of the Scorer functions {(see e.g. Ref.~\cite{Antosiewicz1972})}. 
Such symbol is however not smooth enough ($\HS$ is reduced to the Heaviside step function when the star product---cf.~\eqref{eq:starprod}---is replaced by the ordinary product),
so $\mathcal M_\eta$ cannot be computed within the assumptions of Ref.~\cite{Fagotti2024asymptotic}. The solution to this problem is however already known, it is indeed the Fredholm determinant studied by Tracy and Widom in Ref.~\cite{Tracy1993Level}; $\mathcal M_\infty(z)$, in particular, is the GUE Tracy-Widom distribution $F_2(2^{1/3}z)$.
We have calculated it within a hybrid perturbation theory where $\eta$  is identified as the small parameter, but each order of the expansion is computed using an alternative expansion based on the small parameter $(1-e^{-\eta})$. Though unconventional, this method results in a rapidly converging perturbation series as $\eta\rightarrow 0$ (i.e., $\beta\rightarrow\infty $)---Section~\ref{ss:approx}
\begin{equation}\label{perturbation theory magnetization}
\mathcal M_\eta(z)\sim  \exp\Bigl[-\sum_{n=1}^\infty \eta^n I_n(z)\Bigr]\, .
\end{equation}
The first order of the expansion is
\begin{equation}
I_1(z)=\frac{1}{3\pi} \int_{0}^\infty d y \  \HS(2z+y^{\frac{2}{3}}) ,
\end{equation}
whereas $I_n(z)$ at higher orders is reported in Section~\ref{Sec:order parameter correlations}. Fig.~\ref{fig:mx} shows the profile of the longitudinal magnetization as a function of the rescaled variable at different times compared with the perturbative predictions, for different temperatures of the left reservoir. Fig.~\ref{fig:magnetization_infinite_temperature} highlights the case $\beta=0$, in which the asymptotic behavior is captured by the Tracy-Widom distribution.

Besides the local longitudinal magnetization, we have also studied the entanglement asymmetry~\cite{asymmetry} $\Delta S$ of subsystems, which, arguably, quantifies the order as  effectively as the ``best'' odd observable with support in the subsystem. It is defined for a (reduced) density matrix $\rho_A$ of a (sub)system $A$ as 
\begin{equation}
\Delta S_A=-\mathrm{tr}( \overline{\rho_{A}}\log \overline{\rho_{A}})+\mathrm{tr}(\rho_A\log \rho_A)\, ,
\end{equation}
where $\overline{ \rho_A}$ is the symmetrized reduced density matrix (see section \ref{s:asymmetry}).
For a subsystem $A=\llbracket l,r\rrbracket$ from site $l$ to site $r$ of extent sufficiently larger than the ground-state correlation length, we obtain the following asymptotic behaviour:
\begin{equation}
\Delta S_{(l,r)}=\log 2-H_1(\mathcal M_\eta(z_{r,t}))\, ,
\end{equation}
where $H_1(x)=-\frac{1+x}{2}\log\frac{1+x}{2}-\frac{1-x}{2}\log\frac{1-x}{2}$. 

This expression is remarkable for two reasons: first, it doesn't depend on the left boundary and, second, it depends on the state only through the one-point scaling function. We do not  prove the first property rigorously, but we provide a strong argument in support of it in Section~\ref{s:asymmetry}. 
On the other hand, assuming the first property, the second one is a direct consequence of the following striking feature: 

\begin{description}
\item[{Scaling of local order parameters is universal}] the expectation value of any local odd operator $\bs O_{o,j}$ is characterized by the same scaling function $\mathcal M_\eta(z)$ (with $z=(j-v_{\rm max}t)/|t v''(\bar p)|^{\frac{1}{3}}$).
\end{description}
In particular we have
\begin{equation}\label{general order parameter perturbation theory}
\braket{\bs O_{o,j}}_t=\braket{\textsc{gs}|\bs O_{o}|\textsc{gs}} \mathcal M_\eta \Bigl(\tfrac{j-v_{\rm max}t}{|t v''(\bar p)|^{\frac{1}{3}}}\Bigr)+O(t^{-\frac{1}{3}})\, .
\end{equation}
Thus, the symbol in eq.~\eqref{eq:symboledge} fully characterizes the late-time expectation value of local odd observables in the interfacial region where order depletes. 

This scaling behaviour is not a peculiarity of one-point functions,  indeed also the two point function of the longitudinal spin approaches a function of the rescaled positions
\begin{equation}
\braket{\bs \sigma_j^x\bs \sigma_{n}^x}_{t}=(m^x_{GS})^2\mathcal M_\eta\Bigl(\tfrac{j-v_{\rm max}t}{|t v''(\bar p)|^{\frac{1}{3}}},\tfrac{n-v_{\rm max}t}{|t v''(\bar p)|^{\frac{1}{3}}}\Bigr)+O(t^{-\frac{1}{3}})\, .
\end{equation}
Using a similar perturbative approach as for the one-point function, we express it as 
\begin{equation}
\mathcal M_\eta(z_1,z_2)\sim \exp\left[-\sum_{n=1}^\infty \eta^n I_n(z_1,z_2)\right]\, ,
\end{equation}
and we find
\begin{equation}
I_1(z_1,z_2)=I_1(z_1)-I_1(z_2)
\end{equation}
whereas $I_n(z_1,z_2)$ for higher $n$ can be read from the results reported in Section~\ref{Sec:order parameter correlations}. Again, the two point function of any local odd operator at distance $O(t^{1/3})$ is described by the same scaling function $\mathcal M_\eta(z_1,z_2)$.

By inspecting just the lowest order of the series expansions for the one- and two-point functions we find
\begin{multline}
    \frac{\braket{\bs\sigma^x_j\bs\sigma^x_n}_t}{\braket{\bs\sigma^x_j}_t \braket{\bs\sigma^x_n}_t}=\exp\left[2\eta I_1\Bigl(\tfrac{n-v_{\rm max}t}{|t v''(\bar p)|^{\frac{1}{3}}}\Bigr)+O(\eta^2)\right],
\end{multline}
which is different from $1$ for any ordered pair $(j,n)$ in the interfacial region, for which $\braket{\bs \sigma^x_{j,n}}\neq 0$.
This is a consequence of the fact that {the interfacial region has}
\begin{description}
\item[{Full-range correlations}] the connected 2-point function of the order parameter is a nonzero function of the rescaled position, hence correlations do not decay inside the interfacial region.
\end{description}
This is witnessed by the scaling of the variance of the total longitudinal spin in any subsystem within the interfacial region.
Before discussing it, we introduce two notations that will be extensively used in the rest of the paper:
\begin{itemize}
\item We denote by
\begin{equation}\label{eq:sub}
A_t(z,z')=\llbracket l_{z,t},r_{z',t}\rrbracket
\end{equation}
the subsystem that, at time $t$, is in the right edge comoving frame and consists of the set of integers $l_{z,t},\dots, r_{z',t}$, where
\begin{equation}
\begin{aligned}
\tfrac{l_{z,t}-1-v_{\rm max}t}{|t v''(\bar p)|^{\frac{1}{3}}}< & z\leq \tfrac{l_{z,t}-v_{\rm max}t}{|t v''(\bar p)|^{\frac{1}{3}}}\\
\tfrac{r_{z,t}-v_{\rm max}t}{|t v''(\bar p)|^{\frac{1}{3}}}\leq &z< \tfrac{r_{z,t}+1-v_{\rm max}t}{|t v''(\bar p)|^{\frac{1}{3}}}\, .
\end{aligned}
\end{equation}
\item If $\bs O=\sum_\ell \bs O_\ell$ is an extensive operator, we denote by $\bs O[A_t(z,z')]$ the operator
\begin{equation}
\bs O[A_t(z,z')]=\sum_{l\in \llbracket l_{z,t},r_{z',t}\rrbracket}\bs O_\ell\, .
\end{equation}
\end{itemize}

From the scaling limit of the one- and two-point functions it follows that  the variance of $\bs S^x[A_t(z,z+\zeta )]$ scales as the square of the subsystem's length
\begin{multline}\label{variance perturbation theory}
\lim_{t\rightarrow\infty}\ \frac{\braket{(\bs S^x[A_t(z,z+\zeta )])^2}-\braket{\bs S^x[A_t(z,z+\zeta )]}^2}{|v''(\bar p)t|^{2/3}\zeta^2}=\\
\frac{(m_{\textsc{gs}}^x)^2}{2}\iint\limits_{[\frac{z}{\zeta},\frac{z}{\zeta}+1]\atop y_1<y_2} d^2 y[\mathcal M_\eta(\zeta y_1,\zeta y_2)-\mathcal M_\eta(\zeta y_1)\mathcal M_\eta(\zeta y_2)]=\\
(m_{\textsc{gs}}^x)^2\iint\limits_{[\frac{z}{\zeta},\frac{z}{\zeta}+1]\atop y_1<y_2} d^2 y e^{-\eta I_1(\zeta y_1)}\sinh[\eta I_1(\zeta y_2)]+O(\eta^2) .
\end{multline}
Thus, observables at different rescaled variables $z$ are  strongly correlated notwithstanding being at a distance that approaches infinity as $t\rightarrow\infty$.

In principle, the extraordinary large variance might just account for strong classical correlations reminiscent of the left thermal reservoir (similarly to the effective breakdown of cluster decomposition pointed out in Ref.~\cite{Alba2017Prethermalization}).
We show, however, in Sections~\ref{s:skew} and \ref{sec:physical_int}  that such a slow decay is not limited to classical correlations, indeed { the interfacial region is a}
\begin{description}
\item[{Macroscopic quantum state}] \hspace{0.5 cm }the Wigner-Yanase skew information $I_{\rho_A}(\bs S^x)$ in a generic subsystem $A$ with extent of order $t^{1/3}$ in the interfacial region scales as the square of the subsystem's length $|A|$.
\end{description}
This implies that the size $\mathcal N_A$ of the effective quantum space of $A$ is proportional to $|A|$. Indeed the Wigner-Yanase skew information $I_{\rho_A}(\bs O)$ provides both a lower bound and an upper bound to the quantum Fisher information $F_{\rho_A}(\bs O)$, which is used to bound $\mathcal N_A$ from below~\cite{Frowis2018Macroscopic}.

{In simpler terms, the interfacial region has}
\begin{description}
\item[{Full-range quantum correlations}] also the quantum part of the connected correlation of the order parameter does not decay to zero inside the interfacial region.
\end{description}
This readily follows from  $\frac{1}{2}I_{\rho_A}(\bs S^x)$ being a lower bound to the quantum variance introduced in Ref.~\cite{Frerot2016Quantum}.

For the sake of completeness, Section~\ref{sec:numerical} also exhibits some preliminary numerical data for the (connected) dynamical correlation functions $\braket{\bs\sigma^x_{j}(t\cos\varphi)\bs\sigma_n^x(t\sin\varphi)}_0-\braket{\bs\sigma^x_{j}(t\cos\varphi)}_0\braket{\bs\sigma^x_{n}(t\sin\varphi)}_0$, focusing on the case in which sites $j,n$ correspond to the same scaling variable $z$, but at different times, parametrized by $\varphi\in[\pi/4,\pi/2]$ (i.e. $|t v''(\bar p)|^{\frac{1}{3}} z=j-v_{\rm max}t\cos\varphi=n-v_{\rm max}t\sin\varphi$). 
We provide numerical evidence that also the non-equal time connected correlations do not approach zero in the limit $t\to\infty$ at fixed $\varphi$.

\section{GHD at the edge of the lightcone}\label{s:GHD}
We start with a brief review of the free-fermion techniques useful to deal with Gaussian states that time evolve under the quantum Ising Hamiltonian. We refer the reader to Refs~\cite{Fagotti2016Charges, Alba2021Generalized} for a more comprehensive review.
The Hamiltonian of the transverse-field Ising chain is a quadratic form in the Majorana fermions
\begin{equation}\label{majorana fermions}
    \bs a_{2\ell-1}=\left(\prod_{j=-\infty}^{\ell-1}\bs\sigma^z_j\right)\bs \sigma^{x}_\ell , \qquad \bs a_{2\ell}=\left(\prod_{j=-\infty}^{\ell-1}\bs\sigma^z_j\right)\bs \sigma^{y}_\ell \; ,
\end{equation}
which are self-adjoint operators that satisfy the algebra
$
    \{\bs a_{\ell}, \bs a_n\}=2\delta_{\ell n}\bs I 
$. 
Specifically, the Hamiltonian can be written as $\bs H=\sum_{j,\ell} \bs a_{j} \mathcal{H}_{j,\ell}\bs a_\ell/4$ for some Hermitian antisymmetric matrix $\mathcal{H}$. In the thermodynamic limit $\mathcal{H}$ is a block-Laurent operator generated by a $2$-by-$2$ symbol $h(p)$, that is to say
\begin{equation}
    \begin{pmatrix}
    \mathcal{H}_{2j-1,2\ell-1} & \mathcal{H}_{2j-1,2\ell} \\
      \mathcal{H}_{2j,2\ell-1} & \mathcal{H}_{2j,2\ell} 
    \end{pmatrix}=\int_{-\pi}^\pi \frac{dp}{2\pi} \ h(e^{ip})e^{ip(j-\ell)}
\end{equation}
with $h(e^{ip})=-2\sin (p) \sigma^x-2(h-\cos (p))\sigma^y$. The connection with the standard diagonalisation procedure involving a Bogoliubov transformation in the Fourier space is established by 
the following representation
\begin{equation}
    h(e^{ip})=\varepsilon(p) e^{-i\frac{\theta(p)}{2}\sigma^z}\sigma^y e^{i\frac{\theta(p)}{2}\sigma^z} \;, 
\end{equation}
where $\theta(p)$ is the Bogoliubov angle, given by $e^{i\theta(p)}=-(h-e^{ip})/|h-e^{ip}|$, and $\varepsilon(p)=2\sqrt{1+h^2-2h\cos p}$ is the energy of the quasiparticle excitation with momentum $p$.

It is customary to call a state Gaussian if the expectation value of every  local operator can be expressed in terms of solely  the correlation matrix $\Gamma_{j,\ell}=\delta_{j,\ell}-\braket{\bs a_j \bs a_\ell}$ through the Wick's theorem~\cite{Gaudin1960}. 
If the state is also translationally invariant, its correlation matrix is a block-Laurent operator as well
\begin{equation}
    \begin{pmatrix}
    \Gamma_{2j-1,2\ell-1} & \Gamma_{2j-1,2\ell} \\
      \Gamma_{2j,2\ell-1} & \Gamma_{2j,2\ell} 
    \end{pmatrix}=\int_{-\pi}^\pi \frac{dp}{2\pi} \ \Gamma(e^{ip})e^{ip(j-\ell)}\, ,
\end{equation}
with a $2$-by-$2$ symbol $\Gamma(e^{ip})$~\footnote{If a state is $\kappa$-site shift invariant, the symbol is a $2\kappa$-by-$2\kappa$ matrix function of the momentum.}. This picture can be extended to inhomogeneous  systems by representing the correlation matrix as follows
\begin{equation}\label{correlation matrix elements inhomogenous}
 \Gamma_{2j-2+i,2\ell-2+i'} = \int_{-\pi}^\pi \frac{dp}{2\pi} \ [\Gamma_{\frac{j+\ell}{2}}]_{i, i'} (e^{ip}) e^{ip(j-\ell)} \; ,
\end{equation}
where $i,i'\in\{1,2\}$.
Using the terminology of Ref.~\cite{Fagotti2024asymptotic}, we say that $\Gamma$ is a ``star-Laurent operator'' generated by the symbol $\Gamma_x(e^{i p})$. The symbol of the correlation matrix time evolves according to a Moyal dynamical equation, which is decoupled in the following representation~\cite{Fagotti2020}
\begin{multline}\label{correlation matrix symbol lqss}
  \Gamma_{x,t}(e^{ip})=e^{-i\frac{\theta(p)}{2}\sigma^z}\star\big[4\pi\varrho_{x,t;o}(p) +(4\pi\varrho_{x,t;e}(p)-1)\sigma^y  \\ +4\pi\Psi_{x,t;R}(p)\sigma^z-4\pi\Psi_{x,t;I}(p)\sigma^x \big]\star e^{i\frac{\theta(p)}{2}\sigma^z}  \; ,
\end{multline}
where $\varrho_{x,t}(p)$ is the root density, $\varrho_{e,o}=(\varrho(p)\pm \varrho(-p))/2$ are its even and odd part, respectively, and $\Psi_{x,t}(p)=\Psi_R(p)+i\Psi_I(p)$ is an auxiliary field, which is odd under $p\rightarrow -p$. {The root density is a real field describing the density of excitations, while the auxiliary field is a complex field that captures the creation/annihilation of excitations.} The operation $\star$ is the Moyal product, which is formally defined as follows
\begin{multline}\label{eq:starprod}
    (f\star g)(x,p)  = e^{i\frac{\partial_x\partial_q-\partial_y\partial_p}{2}}f(x,p)g(y,q)\Big|_{y=x,q=p}\\ =\sum_{m,n\in\mathbb{Z}} e^{i(m+n)p} \iint_{-\pi}^{\pi} \frac{d^2q}{(2\pi)^2} \\
    e^{-i(nq_1+mq_2)}f(x-\tfrac{m}{2},q_1)g(x+\tfrac{n}{2},q_2) \ .   
\end{multline}
As anticipated above, the root density and the auxiliary field have independent dynamical equations: the root density time evolves as a Wigner function
\begin{equation}
    i \partial_t \varrho_{x,t}(p)=\varepsilon(p)\star\varrho_{x,t}(p)-\varrho_{x,t}(p)\star\varepsilon(p),
\end{equation}
whereas the auxiliary field satisfies
\begin{equation}\label{eq:Psidyn}
    i \partial_t \Psi_{x,t}(p)=\varepsilon(p)\star \Psi_{x,t}(p)+\Psi_{x,t}(p)\star\varepsilon(-p)\; .
\end{equation} 
In the limit of low inhomogeneity the Moyal equation can be expanded in the order of space derivatives. Keeping the first two non-zero orders gives the third order generalized hydrodynamic equation
\begin{equation}\label{third order generalized hydrodynamics}
    \partial_t\varrho_{x,t}^{(3)}(p)+v(p)\partial_x\varrho^{(3)}_{x,t}(p)=\tfrac{v''(p)}{24}\partial_x^3\varrho^{(3)}_{x,t}(p)\ ,
\end{equation}
where $v(p)=\frac{d\varepsilon(p)}{dp}$ is the velocity of the quasiparticle excitation with momentum $p$. {The superscript $(3)$ is a reminder that $\varrho$ is equal to $\varrho^{(3)}$ only up to subleading contributions---in the limit of infinite time---coming from spatial derivatives higher than the third.} We remark that the third-order correction  is particularly important close to the  edge of the lightcone, where it gives a leading contribution~\cite{Alba2021Generalized}. 
The solution to the third-order GHD equation is readily obtained and reads
\begin{equation}\label{eq:rho3}
\varrho_{x,t}^{(3)}(p)=\int_{-\infty}^\infty \mathrm d y \mathrm{Ai}(y)\varrho^{(3)}_{x-v(p)t+\frac{y}{2}[v''(p)t]^{1/3},0}(p)\; ,
\end{equation}
where $\mathrm{Ai}(y)$ is the Airy function. Since it is reasonable to expect that any nonzero entropy density would kill the ferromagnetic order, we focus our attention to the right edge of the lightcone, where the entropy density approaches zero. 
It is  easily deducible from \eqref{eq:rho3} that the emergent scale around the lightcone is of order $t^{1/3}$.
\begin{figure}
    \includegraphics[width=0.95\linewidth]{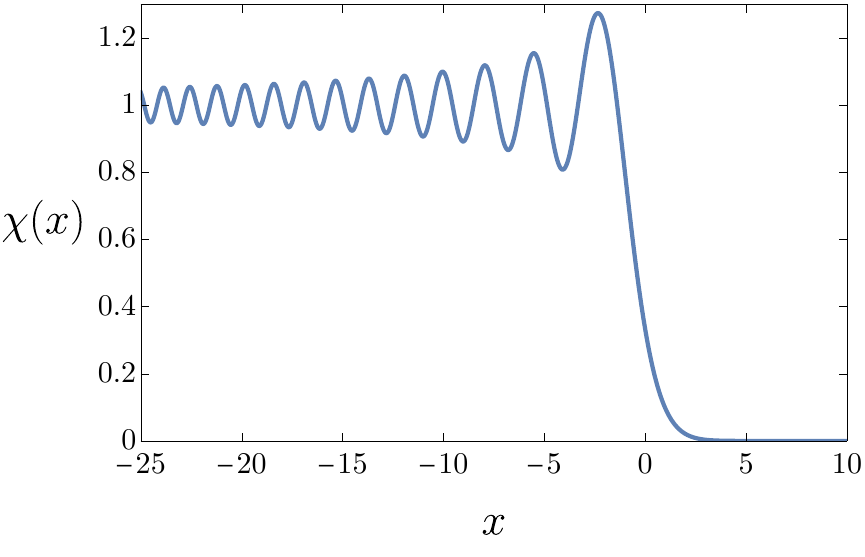}
    \caption{Function $\HS$, appearing in the solution of the third-order GHD equation.}
    \label{fig:chi}
\end{figure}
Let us then rescale the position in the customary way
\begin{equation}\label{eq:scaling}
x=v_{\mathrm{max}}t+|v''(\bar p)|^{1/3} z_{x,t} t^{1/3}\, ,
\end{equation}
where $v(\bar p)=v_{\mathrm{max}}$, and $z_{x,t}$ is the rescaled position. If $\varrho_L(p)$ denotes the root density of the left reservoir, we then have
\begin{multline}
\varrho_{x,t}^{(3)}(p)=\varrho_L(p)\theta_H(v''(p))-\\
\varrho_L(p)\mathrm{sgn}(v''(p))\HS\Bigl(2\tfrac{(v(p)-v_{\mathrm{max}})t^{2/3}-|v''(\bar p)|^{1/3} z_{x,t}}{[v''(p)]^{1/3}}\Bigr)
\end{multline}
where {$\theta_H$ is the Heaviside step function and} $\HS$ is the universal Fermi distribution that Refs~\cite{Bettelheim2011Universal,Dean2018} identified in a degenerate Fermi gas
\begin{equation}
\HS(x)=\int^{\infty}_{x }\!\!\!\mathrm d y \mathrm{Ai}[y] =\pi[\mathrm{Ai}(x)\mathrm{Gi}'(x)-\mathrm{Gi}(x)\mathrm{Ai}'(x)]\; ,
\end{equation}
with $\mathrm{Gi}$ one of the Scorer functions. The function is shown in Fig.~\ref{fig:chi}.  
We are interested in the infinite time limit of local observables with a finite rescaled position $z_{x,t}=z$, i.e., with a position that deviates from the right edge of the lightcone by $O(t^{\frac{1}{3}})$. For given $z$ and fixed momentum $p$ such that $v_{\mathrm{max}}-v(p)$ is nonzero, $\varrho^{(3)}(p)$ approaches zero, which is indicative of the fact that, around the right edge, the state is almost locally equivalent to the ground state. Only excitations with velocity close to the maximal one can be present. Thus, it is convenient to define a rescaled momentum $q_{p,t}$ representing the displacement from the momentum $\bar p$ with the maximal velocity
\begin{equation}
p=\bar p+t^{-1/3} |v''(\bar p)|^{-1/3} q_{p,t}\; .
\end{equation}
At the edge we can find quasiparticles with a finite rescaled momentum $q_{p,t}$, indeed the  limit  of infinite time along the curve $(z_{x,t},q_{p,t})=(z,q)$ is characterised by the following effective root density 
\begin{equation}
\varrho_{x,t}(p)\xrightarrow[(z_{x,t},q_{p,t})=(z,q)]{t\rightarrow\infty}\varrho_L(\bar p)\HS(2z+q^2)\; .
\end{equation}
In the specific case of a thermal left reservoir, we have
\begin{equation}
    \varrho_L(p)=\frac{1}{2\pi}\frac{1}{1+e^{\beta\varepsilon(p)}}\, ,
\end{equation}
hence $\varrho_L(\bar p)=(1-e^{-\eta})/(4\pi)$, where $\eta$ was defined in~\eqref{eta explicit}.
If instead the left hand side is prepared in the ground state of a different noninteracting Hamiltonian, $\varrho_L(\bar p)$ is the root density of the GGE describing the stationary values of local observables sufficiently far to the left of the left edge of the lightcone; specifically, it is given by
\begin{equation}
    \varrho_L(p)=\frac{1}{2\pi}\sin^2(\tfrac{\theta(p)-\theta_0(p)}{2})\, ,
\end{equation}
where $\theta_0(p)$ is the Bogoliubov angle of the prequench Hamiltonian---see, e.g., Ref.~\cite{Fagotti2016Charges} for additional details.  

So far we have only considered the root density contribution, but what about the auxiliary field? 
{
Expanding \eqref{eq:Psidyn} in a reflection symmetric model such as the transverse field Ising chain---in which $\varepsilon(-p)=\varepsilon(p)$--- gives~\cite{Fagotti2020}
\begin{equation}\label{equation auxiliary field asymptotic}
i\partial_t[e^{2i\varepsilon(p)t}\Psi_{x,t}(p)]=-\tfrac{v'(p)}{4}\partial_x^2[e^{2i\varepsilon(p)t}\Psi_{x,t}(p)]+O(\partial_x^4)
\end{equation}
which, at fixed $p$, is a free Schr\"odinger equation, solved by Gaussian wave packets.
}
In a bipartitioning protocol in which the right part is initially prepared in a stationary state (such as in the system under consideration), the auxiliary field {is zero on the right hand side, and thus satisfies the initial condition $\Psi_{x,0}(p)=\theta_H(-x)\Psi_L(p)$, where $\Psi_L(p)$ is the auxiliary field of the left part. Solving Eq.~\eqref{equation auxiliary field asymptotic} then gives
$$
\Psi_{x,t}(p)\sim  \pi e^{-2i\varepsilon(p)t}\Psi_L(p)\mathrm{erfc}\Bigl(e^{-i\frac{\pi}{4}\mathrm{sgn}(v'(p))}\tfrac{x}{\sqrt{|v'(p)t|}}\Bigr),
$$
where $\mathrm{erfc}$ is the complementary error function. At finite times the auxiliary field is characterized by rapid oscillations and an absolute value that scales as $\sim \frac{\sqrt{|v'(p)|t}}{x}$.
Along any ray with a finite slope strictly within the lightcone, the auxiliary field is characterized by an absolute value that behaves as $\sim t^{-1/2}$, while around the edge of the lightcone, along the curve $(z_{x,t},q_{p,t})=(z,q)$, the behavior is $\sim t^{-2/3}$. Hence, the auxiliary field gives a negligible contribution around the edge of the lightcone and the root density provides a complete description.}
This observation {also} allows us to regard nonequilibrium preparations  of the left part of the chain (resulting from global quenches) as stationary preparations.

In order to reconstruct the symbol of the correlation matrix we should apply the canonical transformation $(x,p)\rightarrow (z,q)$ to the Bogoliubov angle and to the star product. The star product is unchanged
\begin{equation}
(f\star g)(z_{x,t},q_{p,t})  = e^{i\frac{\partial_z\partial_\kappa-\partial_\zeta\partial_q}{2}}f(z,q)g(\zeta,\kappa)\Big|_{\zeta=z\atop\kappa=q}\; ,
\end{equation}
where, on the right hand side, the subscripts in $z$ and $q$ are understood.  For a momentum with velocity close to its maximum, the Bogoliubov angle can be expanded as follows
\begin{equation}
\theta(p)=\theta(\bar p)+\theta'(\bar p)t^{-1/3} |v''(\bar p)|^{-1/3} q_{p,t} +O(t^{-2/3})\, .
\end{equation}
Consequently, every derivative with respect to the rescaled momentum coming from the Moyal product with the Bogoliubov phase in \eqref{correlation matrix symbol lqss} gives an $O(t^{-1/3})$ contribution, which can be neglected in the limit $t\rightarrow\infty$. The Moyal product is therefore reduced  to a conventional product, and we can express the symbol of the correlation matrix in the following suggestive way
\begin{multline}\label{eq:corr_mat}
\Gamma_{x }(e^{ip})\sim
\tfrac{1-e^{-\eta}}{2}
\Bigl[\HS(2z_{x}+q_{p}^2)-\HS(2z_{x}+q_{-p}^2)\Bigr]\mathrm I+\\
\Big[1-\tfrac{1-e^{-\eta}}{2}\big(\HS(2z_{x}+q_{p}^2)+\HS(2z_{x}+q_{-p}^2)\big)\Big]\Gamma_{\textsc{gs}}(e^{ip})\; , 
\end{multline}
where the time dependence is understood and $\Gamma_{\textsc{gs}}(e^{i p})=-\sigma^y e^{i\theta(p)\sigma^z}$ is the symbol of the ground state's correlation matrix. 

We remark that, if we take the infinite time limit along the curve $(z_{x,t},q_{\pm p,t})=(z,\pm q)$, the symbol of the correlation matrix commutes with the symbol of the Laurent operator associated with the Hamiltonian. This confirms that, at the level of the correlation matrix, the infinite time limit along the curve $z_{x,t}=z$ is locally described by a stationary state.

\section{Order parameter correlations}\label{Sec:order parameter correlations}

While local observables that are even under spin flip can be computed using the Wick's theorem based on the homogeneous correlation matrix~\eqref{eq:corr_mat} with $z_{x,t}=z$, odd observables have a semilocal fermionic representation (cf. \eqref{majorana fermions}) and depend on the half-infinite correlation matrix whose corner corresponds to the position of the operator. For example, let us consider the typical order parameter of the quantum Ising model, i.e., the  local longitudinal magnetization. 
By the Lieb-Robinson bounds~\cite{Lieb1972The}, we can compute the one-point function of the order parameter from its two-point function with an operator located far to the right of the lightcone (see also Refs~\cite{Bravyi2006Lieb,Bertini2016Determination})
\begin{equation}
    \braket{\bs\sigma_{\ell}^x}_t=\lim_{n\to\infty}\frac{\braket{\bs\sigma_{\ell}^x\bs\sigma_{n}^x}_t}{\braket{\bs\sigma^x}_{\textsc{GS}}}\; .
\end{equation}
The two point function is proportional to the Pfaffian of the matrix obtained by removing
the first row and column from the  correlation matrix of the subsystem $(\ell,\infty)$, which we call $\Gamma^{xx}$. Specifically, we have 
\begin{equation}\label{two point correlator pfaffian}
\braket{\bs\sigma_\ell^x\bs\sigma_n^x}=i^{n-\ell}\mathrm{Pf}[\Gamma^{xx}_{(\ell, n)}]\, ,
\end{equation} 
where $\Gamma^{xx}_{(\ell, n)}$ is the finite section of $\Gamma^{xx}$ from site $2\ell-1$ to $2n$.
For a ``star block-Toeplitz matrix'' (an inhomogeneous semi-infinite matrix, as dubbed in Ref.~\cite{Fagotti2024asymptotic}) generated by a 2-by-2 symbol such as \eqref{eq:corr_mat}, removing the first row and column is equivalent to transforming the symbol as follows
\begin{multline}\label{eq:Gammaxx}
\Gamma_{x}(e^{ip})\equiv\begin{pmatrix}
{}[\Gamma_{x}(e^{ip})]_{11}&[\Gamma_{x}(e^{ip})]_{12}\\
[\Gamma_{x}(e^{ip})]_{21}&[\Gamma_{x}(e^{ip})]_{22}
\end{pmatrix}\mapsto \\
\begin{pmatrix}
{}[\Gamma_{x}(e^{ip})]_{22}&e^{ip}[\Gamma_{x+\frac{1}{2}}(e^{ip})]_{21}\\
e^{-ip}[\Gamma_{x+\frac{1}{2}}(e^{ip})]_{12}&[\Gamma_{x+1}(e^{ip})]_{11}
\end{pmatrix}\equiv \Gamma_{x}^{xx}(e^{ip})\, .
\end{multline}
Importantly,  $\Gamma_{x,t}(e^{ip})$ in \eqref{eq:corr_mat}, and hence  $\Gamma^{xx}_{x,t}(e^{ip})$ in \eqref{eq:Gammaxx}, is different from the symbol of the ground state correlation matrix only for an order one phase-space region (i.e., for $x$ in the interfacial region, which is $\sim t^{1/3}$, and $p$  close to the momentum with the maximal velocity, which is $\sim t^{-1/3}$). A $t^{-1/3}$ correction confined in the same region  does not affect the asymptotic behaviour of the Pfaffian, hence we can confuse positions $x+\frac{1}{2}$ and $x+1$ with $x$ and simplify the transformation in 
$\Gamma^{xx}_{x,t}(e^{ip})\sim e^{i\frac{p}{2}\sigma^z}\sigma^x \Gamma_{x,t}(e^{ip})\sigma^xe^{-i\frac{p}{2}\sigma^z}$, where we also used $\mathrm{tr}[\Gamma_{x,t}(e^{ip})\sigma^z]=0$. Thus we have
\begin{multline}
\Gamma^{xx}_{x}(e^{ip})\sim \tfrac{1-e^{-\eta}}{2}\Bigl[\HS(2z_{x}+q_{p}^2)-\HS(2z_{x}+q_{-p}^2)\Bigr]\mathrm I+\\
\Big[1-\tfrac{1-e^{-\eta}}{2}\big(\HS(2z_{x}+q_{p}^2)+\HS(2z_{x}+q_{-p}^2)\big)\Big]\Gamma^{xx}_{\textsc{gs}}(e^{ip})\; .
\end{multline}

In our setting the one-point function has the sign of the order parameter in the ground state prepared on the right hand side of the junction, which we agreed to be positive, hence we can write
\begin{equation}
\lim_{n\rightarrow\infty}\frac{\braket{\bs\sigma_\ell^x\bs\sigma_n^x}_t}{(m^x_{\textsc{gs}})^2}=\exp\left[\frac{1}{2}\mathrm{tr}\left(\log ((\Gamma^{xx}_{\textsc{gs}(\ell,\infty)})^{-1}\Gamma^{xx}_{(\ell,\infty)})\right)\right]
\end{equation}
where we isolated the ground state contribution. The symbol of the product of two Toeplitz operators  is not the product of their symbols; and a similar result holds true for inhomogeneous matrices. The error in carrying  products and inverse at the level of symbols for an operator with a smooth symbol without zeros, such as  $\Gamma^{xx}_{\textsc{gs}(\ell,\infty)}$, is however quasilocalised around the corner of the matrix, which is negligible in the scaling limit under consideration. 
Thus we have
\begin{multline}\label{eq:xxinW}
\lim_{n\rightarrow\infty}\frac{\braket{\bs\sigma_\ell^x\bs\sigma_n^x}_t}{(m^x_{\textsc{gs}})^2}=\\
e^{\frac{1}{2}\mathrm{tr}\bigl(\log \bigl(\mathrm I-W^{(+)}_{(\ell,\infty)}\frac{\mathrm I+\Gamma^{xx}_{\textsc{gs}(\ell,\infty)}}{2}-W^{(-)}_{(\ell,\infty)}\frac{\mathrm I-\Gamma^{xx}_{\textsc{gs}(\ell,\infty)}}{2}\bigr)\bigr)}
\end{multline}
where $W^{\pm}$ are generated by the scalar symbols
\begin{equation}
W^{(\pm)}_x(e^{ip})=(1-e^{-\eta})\HS(2z_{x,t}+q_{\mp p,t}^2)\, .
\end{equation}
We can use the same argument as before to infer that $\frac{\mathrm I\pm\Gamma^{xx}_{\textsc{gs}(\ell,\infty)}}{2}$ can be treated as projectors
\begin{equation}
\frac{\mathrm I+s\Gamma^{xx}_{\textsc{gs}(\ell,\infty)}}{2} \frac{\mathrm I+s'\Gamma^{xx}_{\textsc{gs}(\ell,\infty)}}{2}{\sim} \delta_{s s'}\frac{\mathrm I+s\Gamma^{xx}_{\textsc{gs}(\ell,\infty)}}{2}\; ,
\end{equation}
with $s,s'=\pm 1$. Since $W^{+}$ and $W^{-}$ are equivalent and the scaling variables are such that sums over indices can be turned into integrals up to corrections $O(t^{-1/3})$, we can finally express the local longitudinal magnetization as a Freedholm determinant 
\begin{equation}\label{eq:sxdet}
\braket{\bs\sigma_{\ell}^x}_t\sim m^x_{\textsc{gs}} \det\left|\mathrm I-(1-e^{-\eta}) \hat n_{z_{\ell,t}}\right|\; ,
\end{equation}
where $\hat n_{z_0}(z_1,z_2)$ is the Airy kernel
\begin{multline}
\hat n_{z_0}(z_1,z_2)= \int_{-\infty}^\infty \frac{d q}{2\pi} \   \HS\left(2z_0+z_1+z_2+q^2\right)e^{iq(z_1-z_2)}\\
=\frac{1}{z_1-z_2}\Big(\mathrm{Ai}[2^{\tfrac{1}{3}}(z_0+z_1)]\mathrm{Ai}'[2^{\tfrac{1}{3}}(z_0+z_2)]\\-\mathrm{Ai}'[2^{\tfrac{1}{3}}(z_0+z_1)]\mathrm{Ai}[2^{\tfrac{1}{3}}(z_0+z_2)]\Big)
\end{multline}

and $z_1,z_2\geq 0$. 

The two-point function has an analogous expression with respect to a finite section of the operator $\hat n_{z_0}$
\begin{equation}\label{eq:sxsxdet}
\braket{\bs\sigma_{\ell}^x\bs\sigma_{n}^x}_t\sim (m^x_{\textsc{gs}})^2 \det\left|\mathrm I-(1-e^{-\eta}) \hat n_{z_{\ell,t},z_{n,t}}\right|
\end{equation}
where $\hat n_{z_{\ell,t},z_{n,t}}(z_1,z_2)=\hat n_{z_{\ell,t}}(z_1,z_2)$ for $z_{\ell,t}<z_1,z_2<z_{n,t}$.

\subsection{Universality of the scaling functions}
Even if we derived  \eqref{eq:sxdet} and \eqref{eq:sxsxdet} focusing on the local longitudinal spin, we point out that most of the  changes associated with replacing $\bs\sigma_\ell^x$ by another odd local observable are subleading. The reason is that the expectation value of any  odd local observable that can be written as the product of Pauli matrices is  represented by a Pfaffian of a matrix that is identical to $\Gamma^{(xx)}_{(\ell, n)}$ except for a finite number of rows and columns. The latter affects the asymptotic expression only through the expectation value of the operator in the ground state. Specifically we find
\begin{equation}
\begin{aligned}
\braket{\bs O_x}_t\sim &\braket{\bs O}_{\textsc{gs}} \det\left|\mathrm I-(1-e^{-\eta}) \hat n_{z_{x,t}}\right| \label{universal relation O one point}\\
\braket{\bs O^{\phantom\prime}_x\bs O'_y}_t\sim &\braket{\bs O}_{\textsc{gs}}\braket{\bs O'}_{\textsc{gs}} \det\left|\mathrm I-(1-e^{-\eta}) \hat n_{z_{x,t},z_{y,t}}\right|\; ,
\end{aligned}
\end{equation}
where $\bs O,\bs O'$ are odd local observables at a finite rescaled distance from each other.
Following the same reasoning, we also argue
\begin{equation}
\braket{\bs O_x\bs E_y}_t\sim \braket{\bs O}_{\textsc{gs}}\braket{\bs E}_{\textsc{gs}} \det\left|\mathrm I-(1-e^{-\eta}) \hat n_{z_{x,t}}\right|\; ,
\end{equation}
where $\bs E$ is an even local observable at a finite rescaled distance from $\bs O$ ($y$ can be either greater or less than $x$). At a finite rescaled distance, instead,  local observables that are even under spin flip cluster and approach the ground-state expectation values. 

Incidentally, a similar reasoning applies to non-equal time two-point functions. In particular, we argue
\begin{equation}\label{universal relation dynamical}
    \frac{\braket{\bs O_x(t\cos\phi)\bs O_y(t\sin\phi)}_0}{\braket{\bs O'_x(t\cos\phi)\bs O'_y(t\sin\phi)}_0}\sim \frac{\braket{\bs O}_{\textsc{gs}}^2}{\braket{\bs O'}_{\textsc{gs}}^2} \; ,
\end{equation}
where $0$ stands for the initial state{, and $O,O'$ are odd local observables as in Eq.~\ref{universal relation O one point}}. Because of \eqref{universal relation O one point}, the same result applies also to the connected part of the correlation function.

\begin{figure}[t]
    \centering
  \includegraphics[width=1\linewidth]{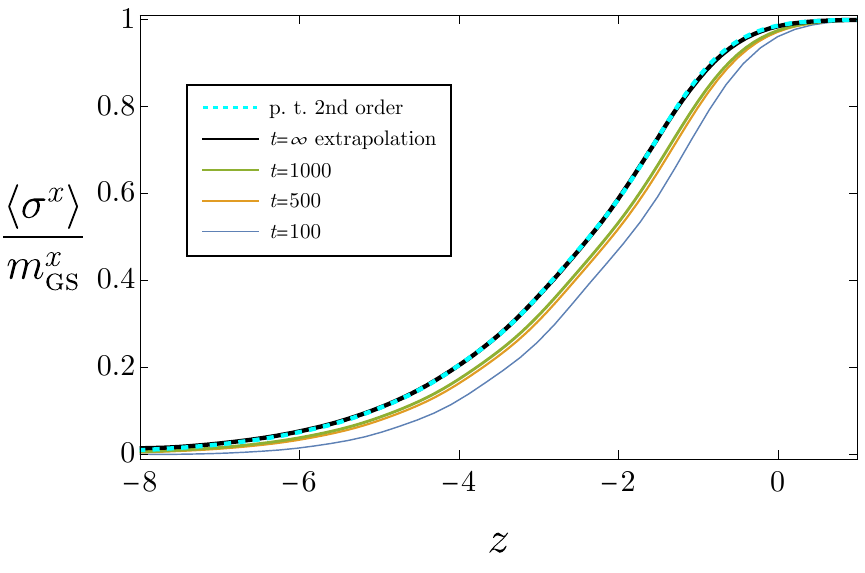}
    \caption{The same as in Fig.~\ref{fig:mx}, except that the left part of the system is prepared in the ground state of the model with $h=2$ (paramagnetic phase), rather than in an equilibrium state.}
    \label{fig:paramagnetic}
\end{figure}

The second sign of universality that we would like to point out comes from having characterised the left reservoir through a single parameter, $\eta$. Such a simplification is particularly strong if we take into account that our derivation is not specific to thermal reservoirs. Indeed, the left part of the initial state is characterised by a single parameter even if it is prepared in a generalised Gibbs ensemble, or 
in a nonequilibrium state. For instance, Fig.~\ref{fig:paramagnetic} shows  an example in which the left hand side of the chain is prepared in the ground state of the Ising model in a different magnetic field.   

The third  aspect of universality is that the scaling functions are also stable under localised perturbations. This can be shown as follows. Let $\bs B^{\textsc{l}}=(\bs B^{\textsc{l}})^\dag$ be the boundary edge mode of the model, which is an odd conserved involution with support quasilocalised around the left boundary of the right semi-infinite chain (see, e.g., Ref.~\cite{Fagotti2016Local}) and satisfying $\bs B^{\textsc{l}} \ket{\textsc{gs}}=\ket{\textsc{gs}}$. We consider the effect of applying to the initial state a unitary transformation $\bs U$, with support quasilocalised around the junction. 
We can always decompose $\bs U$ as follows:
\begin{equation}\label{perturbation decomposition even odd}
\bs U=\bs W_{+}+\bs W_{-}\bs B^{\textsc{l}}\, ,
\end{equation}
where $\bs W_{\pm}$ are even under spin flip. Since the symmetry-breaking ground state is an eigenstate of  $\bs B^{\textsc{l}}$, $\bs U$ can be replaced by  $\bs W=\bs W_{+}+\bs W_{-}$.
Thus we have
\begin{multline}
\mathrm{tr}(e^{-i \bs H t}\bs U[\rho_\beta\otimes\ket{\textsc{gs},+}\bra{\textsc{gs},+}]\bs U^\dag e^{i \bs H t}\bs \sigma_\ell^x\bs \sigma_n^x)=\\
\mathrm{tr}(e^{-i \bs H t}\bs W[\rho_\beta\otimes\rho_{\infty}]\bs W^\dag e^{i \bs H t}\bs \sigma_\ell^x\bs \sigma_n^x)\, ,
\end{multline}
where we denoted the Gibbs ensemble with inverse temperature $\beta$ by $\rho_\beta$ and we allowed ourselves to replace $\ket{\textsc{gs},+}\bra{\textsc{gs},+}$ by $\rho_{\infty}$ because $\bs W$ is even under spin flip. 
Being quasilocal and even, $\bs W$ can be approximated by a finite sum of Gaussians with support quasilocalised around the junction. The contribution from each of these Gaussians is captured by a pair of root density and  auxiliary field that differs from the one without $\bs U$ only in the neighborhoods of  the junction. As discussed in Ref.~\cite{Alba2021Generalized}, such perturbations affect the edge of the lightcone at $O(t^{-2/3})$ and are therefore negligible in the scaling limit that we consider. 

The fourth aspect of universality that we would like to emphasize is that the scaling functions are not specific to the transverse-field Ising chain. We chose the model only for its simplicity and notoriety, but the same formulas apply, in particular, to any quantum spin chain that can be mapped to free fermions by a Jordan-Wigner transformation, such as the quantum XY model, provided to be in a phase with spontaneous symmetry breaking. 
Minor changes are expected only when there is more that one mode associated with the maximal velocity, such as in the quantum XY model without external field.

\subsection{Approximation of Fredholm determinants}\label{ss:approx}
\begin{figure}
    \centering
    \includegraphics[width=0.95\linewidth]{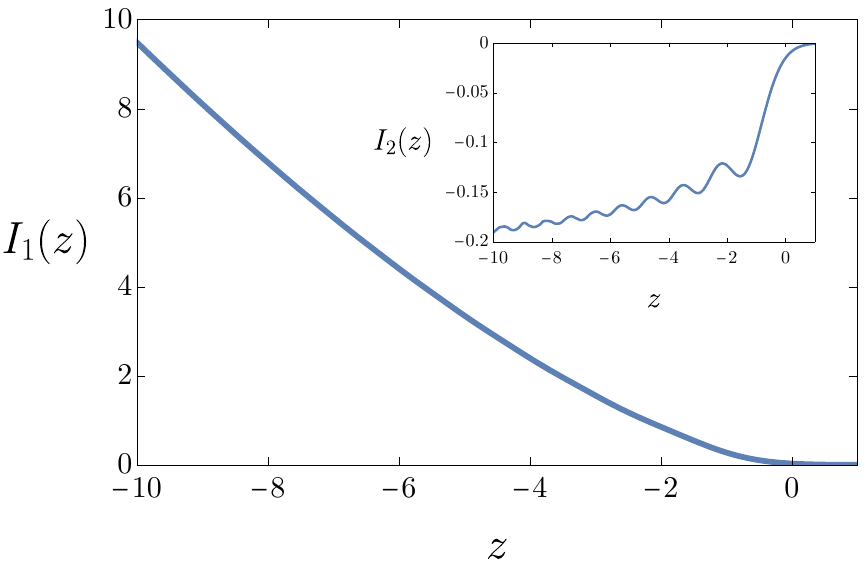}
    \caption{Functions $I_1$ (main plot) and $I_2$ (inset), appearing in the expansion of the Fredholm determinant.}
    \label{fig:i1i2}
\end{figure}

The limit of low temperature (or of small quench, in the sense of Ref.~\cite{Calabrese2012QuantumI}) corresponds to the limit $\eta\rightarrow 0$, and $1-e^{-\eta}$ presents itself as the natural small parameter in the  trace expansion of the Fredholm determinant. On the other hand, the logarithm of the determinant is mainly determined by the eigenvalues of $\hat n_{z}$ that are close to $1$ (note that $\hat n_{-\infty}$ is an involution), for which $\eta$ is the natural parameter of the expansion. Thus, we consider the set of polynomials $p_n$ generated by 
\begin{equation}
-\log(1-(1-e^{-w})x)=\sum_{n=1}^\infty w^n p_n(x)
\end{equation}
and define
\begin{equation}
\begin{aligned}
I_n(z)=&\mathrm{tr}(p_n(\hat n_{z}))\\
I_n(z_1,z_2)=&\mathrm{tr}(p_n(\hat n_{z_1,z_2}))\, .
\end{aligned}
\end{equation}
We mention that, for $n>1$, such polynomials satisfy $p_n(x)=(-1)^n p_n(1-x)$ and report the first of them
\begin{equation}
\begin{aligned}
p_1(x)=&x\\
p_2(x)=&\tfrac{1}{2}(x^2-x)\\
p_3(x)=&\tfrac{1}{6}(2 x^3 - 3 x^2 +x)\\
p_4(x)=&\tfrac{1}{24}(6x^4-12 x^3 +7 x^2-x)\, .
\end{aligned}
\end{equation}
The  one- and two-point scaling functions are then given by
\begin{equation}
\begin{aligned}
\det\left|\mathrm I-(1-e^{-\eta}) \hat n_{z}\right|=&\exp\Bigl(-\sum_{n=1}^\infty \eta^n I_n(z)\Bigr)\\
\det\left|\mathrm I-(1-e^{-\eta}) \hat n_{z_1,z_2}\right|=&\exp\Bigl(-\sum_{n=1}^\infty \eta^n I_n(z_1,z_2)\Bigr)
\end{aligned}
\end{equation}
We observe that these series are rapidly convergent for generic $z$ even at rather high temperature, therefore we can use the finite sums of the first few terms of the series as excellent approximations of the asymptotic behaviour. 
Since the contribution from $I_2(z)$ starts being visible for $\beta\lesssim 1$, we  report an integral representation  that can be evaluated numerically without special precautions
\begin{multline}
   I_2(z)=\frac{1}{2}\Bigl[\iint_{-\infty}^\infty \frac{d q_1 dq_2 }{(2\pi)^2}\iint_{z}^{
  \infty}d y_1 d y_2 e^{i(q_1-q_2)(y_1-y_{2})}\\
   \chi\left(y_1+y_{2}+q_1^2 \right)\chi\left(y_1+y_{2}+q_2^2 \right)\Bigr] -\frac{1}{2}I_1(z)   \ .
\end{multline}
Functions $I_1$ and $I_2$ are shown in Fig.~\ref{fig:i1i2}.

We remind the reader that in the opposite limit $\eta\rightarrow\infty$, $\mathcal M_\eta(z)$ approaches the GUE Tracy-Widom distribution (Fig.~\ref{fig:magnetization_infinite_temperature}).

\subsection{Numerical checks}\label{sec:numerical}

\begin{figure}[t]
    \centering
  \includegraphics[width=1\linewidth]{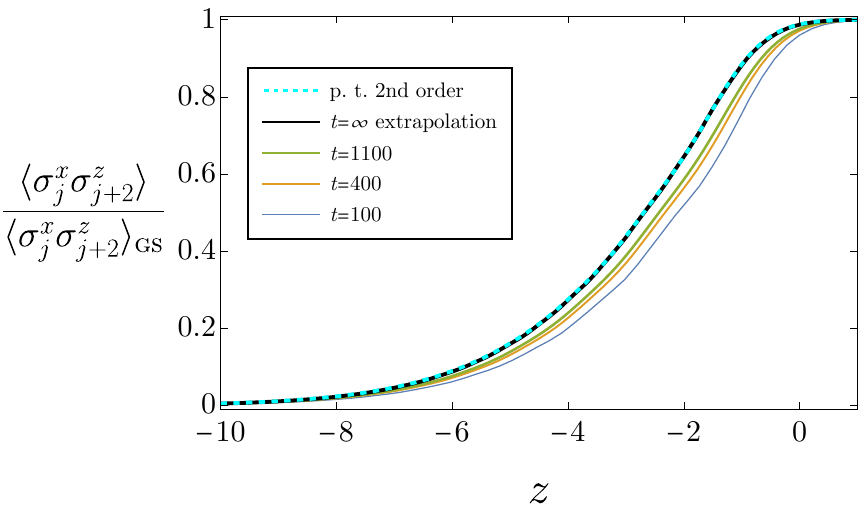}
    \caption{{The same as in Fig.~\ref{fig:mx}(b), but for the order parameter $ \braket{\bs\sigma_j^x\bs\sigma_{j+2}^z}$. For comparison, the ground state values are $m^x_{\textsc{gs}}\approx 0.965$, $\braket{\bs\sigma_j^x\bs\sigma_{j+2}^z}_{\textsc{gs}}\approx 0.249$.}
    }
    \label{fig:XZ}
\end{figure}

\begin{figure}[t]
    \centering
    \includegraphics[width=1.05\linewidth]{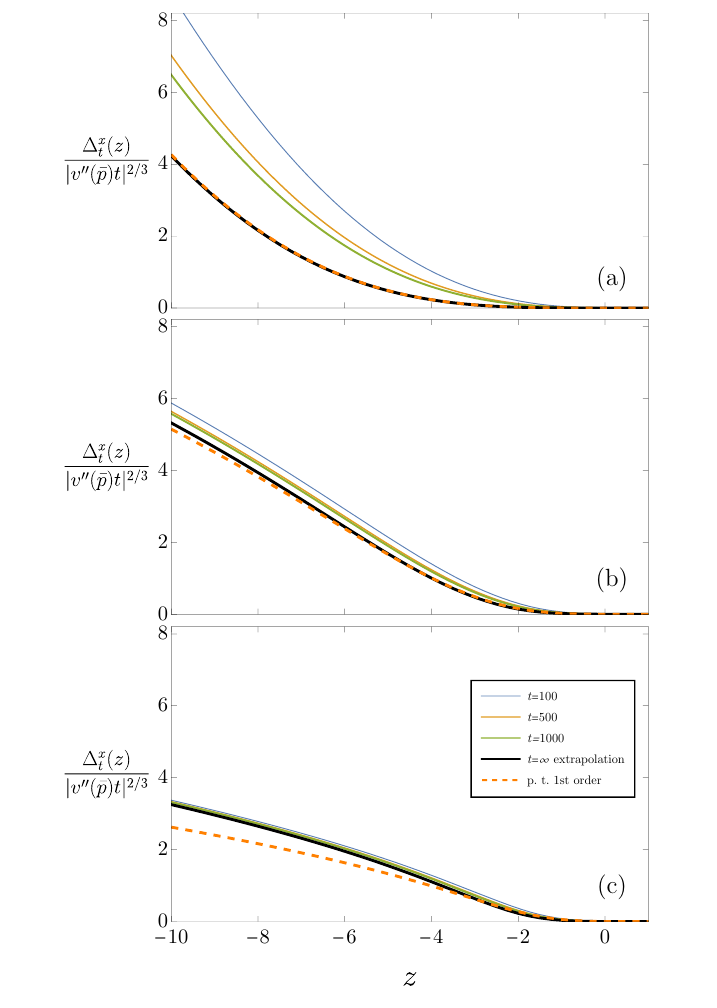}
    \caption{Difference between the variance of $\bs S^x=\frac{1}{2}\sum_{\ell\geq j}\bs \sigma_{\ell}^x$ after the quench and the same variance in the ground state of the model, as a function of the rescaled position $z=(j-v_{\rm max}t)/|t v''(\bar p)|^{\frac{1}{3}}$, at different times $t$ with  a) $\beta=2$, b) $\beta=0.75$, c) $\beta=0.25$. The other parameters are the same as in Fig.~\ref{fig:mx}.
    }
    \label{fig:variance}
\end{figure}

\begin{figure}
    \centering
    \includegraphics[width=0.95\linewidth]{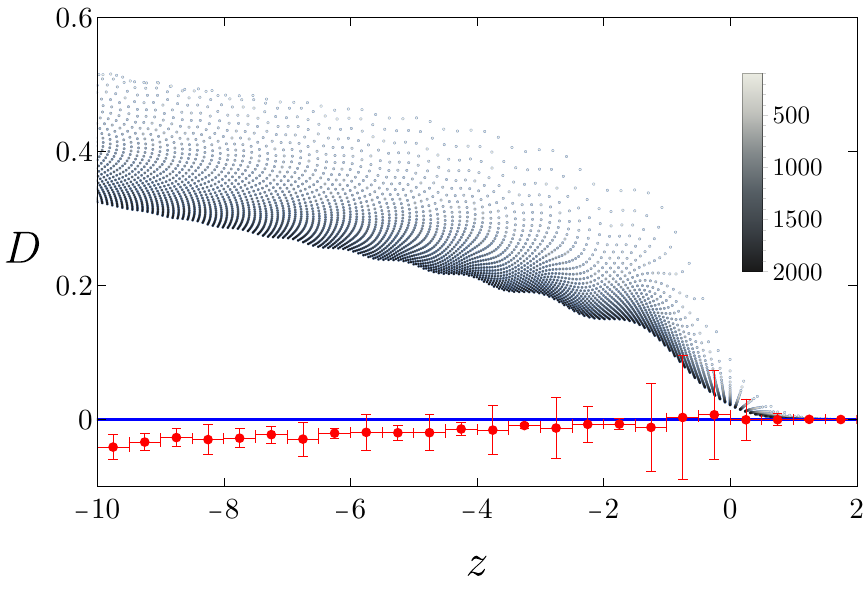}
    \caption{The quantity $D(z)$, defined in Eq.~\eqref{difference defect}, with $\alpha_1=2\alpha_2=1$, as a function of the rescaled position $z=(j-v_{\rm max}t)/|t v''(\bar p)|^{\frac{1}{3}}$. Different times $t$ are indicated by different shades of gray (legend). The extrapolation to $t\to\infty$ {(see Section~\ref{sec:numerical} for details)} is represented by red points, and the zero value by a solid blue line. The parameters of the model are $h=0.5, \beta=0.75$.}
    \label{fig:defect}
\end{figure}

We have compared our analytical predictions against numerical data considering time evolution in a finite-size system with open boundary conditions. We employed standard methods based on the mapping to free fermions.

\paragraph{One-point function.}
In Fig.~\ref{fig:mx} we report the time evolution of the magnetization $\braket{\bs\sigma_j^x}$ as a function of the rescaled variable $z_{j,t}=\tfrac{j-v_{\rm max}t}{|t v''(\bar p)|^{\frac{1}{3}}}$, for several different temperatures of the left thermal reservoir. Assuming corrections of the form $a t^{-1/3}+bt^{-2/3}$, for some coefficients $a,b$, we extrapolate the results to $t\to\infty$. The extrapolated limit is in excellent agreement with prediction~\eqref{perturbation theory magnetization}, which we approximate by replacing the series with a finite sum of a sufficiently large number of terms. Since the parameter of the  perturbation theory increases with temperature, for a left reservoir at higher temperature, a higher number of terms is required.

In Fig.~\ref{fig:paramagnetic} we present an analogous study where the left part of the system is prepared out of equilibrium in the ground state of the Ising model with a different magnetic field within the paramagnetic phase. Again, the agreement with the prediction is excellent. In Fig.~\ref{fig:XZ} we show the alternative order parameter $\braket{\bs \sigma_j^x\bs \sigma_{j+2}^z}$, confirming the universality of the scaling function~\eqref{general order parameter perturbation theory}, in that it is not specific to the longitudinal magnetization. 

\paragraph{Two-point function.}
We find excellent agreement also with our predictions for the order parameter two-point function. For instance, Fig.~\ref{fig:variance} shows the {variance} of the  longitudinal component of the total spin in the semi-infinite subsystem $A_t(z,\infty)$---eq.~\eqref{eq:sub}.
Specifically, we consider the difference
\begin{equation}\label{variance difference}
    \Delta^{x}_t(z)\equiv \mathrm{Var}_t(\bs S^x[A_t(z,\infty)])-\mathrm{Var}_{\GS}(\bs S^x[A_t(z,\infty)]) \ ,
\end{equation}
where $\mathrm{Var}_t(O)\equiv\braket{\bs O^2}_t-\braket{\bs O}_t^2$ denotes the variance at time $t$
and $\mathrm{Var}_\GS$ the variance in the ground state. 
We have subtracted the latter so as to remove the divergency of the variance{, coming from the limit of infinite length of the subsystem outside the lightcone}. 
Note that replacing $A_t(z,\infty)$ by $A_t(z,z_r)$ with $z_{\rm{r}}>0 $ would introduce just exponentially small corrections in $z_{\rm{r}}$ (the state becomes indistinguishable from the ground state at a  positive large enough rescaled position). Thus, the effective size of the subsystem is not infinite but rather scales as $|v''(\bar p) t|^{1/3}$. Since the ground state variance would then scale as $|v''(\bar p) t|^{1/3}$, by computing $\Delta^{x}_{t}(z)/|v''(\bar p)t|^{2/3}$  we are accessing the asymptotics  of the variance in a subsystem including the edge of the lightcone.  
In particular, we can extract the first order of the perturbation theory from eq.~\eqref{variance perturbation theory}
\begin{multline}\label{variance to the right edge perturbation theory}
\lim_{t\rightarrow\infty}\frac{\Delta^{x}_t(z)}{|v''(\bar p)t|^{2/3}}=\\
(m_{\textsc{gs}}^x)^2\int_z^\infty d y_1\int_{y_1}^\infty d y_2 \ e^{-\eta I_1(y_1)}\sinh[\eta I_1(y_2)] \; .
\end{multline}
We have compared this prediction with the extrapolation of our data to $t\to\infty$. As for the one-point function, the lower the temperature of the left reservoir, the better the agreement. To correctly capture the asymptotics at higher temperatures would require retaining higher orders of the perturbation theory.

\paragraph{Stability.}
We have tested the resilience of the late-time behaviour of the one-point function to localized perturbations in the initial state{, i.e. that  $\braket{ W\bs\sigma_j^x(t) W}_{0}\sim\braket{\bs\sigma_j^x(t) }_{0}$ in the limit $t\to\infty$ for perturbations $W$ localized around the junction, where $0$ stands for the initial state. As follows from the discussion around Eq.~\eqref{perturbation decomposition even odd}, it is sufficient to check the resilience against perturbations $W$ that are even under spin flip. Such perturbations can be approximated by a sum of Gaussians $\rho_1,\rho_2,\ldots$ with support localised around the junction
\begin{equation}\label{sum perturbation gaussians}
\braket{ W\bs\sigma_j^x(t) W}_{0}=\sum_{\ell,n} \braket{\rho_\ell \sigma_j^x(t)\rho_n}    \ .
\end{equation}
In fact, our numerical analysis suggests that a stronger condition is met:
each term in the sum in Eq.~\ref{sum perturbation gaussians} satisfies
\begin{equation}
 \braket{\bs\sigma_j^x(t) }_{0}\sim\frac{\braket{\rho\bs\sigma_j^x(t)\rho'}_0}{\braket{\rho\rho'}_0} \ ,  
\end{equation}
in the limit $t\to\infty$, for any pair of Gaussians $\rho,\rho'$ localised around the junction. In Fig.~\ref{fig:defect} we report the data for Gaussians $\rho=e^{i\alpha_1\bs\sigma_1^z}$, $\rho'=e^{i\alpha_2\bs\sigma_1^y\bs\sigma_2^y}$, with $\alpha_1=2\alpha_2=1$. There, we have quantified the deviation as
\begin{equation}\label{difference defect}
   D(z_{j,t})\equiv\frac{\braket{\bs\sigma_j^x(t)}_{0}-\frac{\braket{e^{i\alpha_1\bs\sigma_1^z}\bs\sigma_j^x(t) e^{i\alpha_2\bs\sigma_1^y\bs\sigma_2^y}}_{0}}{\braket{e^{i\alpha_1\bs\sigma_1^z}e^{i\alpha_2\bs\sigma_1^y\bs\sigma_2^y}}_{0}}}{m^x_{\GS}\mathcal M(z_{j,t})} \; .
\end{equation}
We} have grouped the points in bins and made a fit with corrections $t^{-1/3}$. We have attached an  error $t^{-2/3}$ to the data so as to effectively capture the subleading orders of the asymptotic expansion in the limit of infinite time. The error on the rescaled position is artificial and represents the width of the bins used. The error bars on $D$ correspond to the $95\%$ confidence interval on the estimation of the parameters of the fit, which should be regarded just as a suggestive estimation.
Considering that the larger $-z$ and the larger are expected to be the asymptotic corrections, 
we conclude that the extrapolation to $t\to\infty$ is consistent with the prediction. This, in turn, supports the stability of the result under localised perturbations in the initial state. 

\begin{figure}
    \centering
    \includegraphics[width=0.95\linewidth]{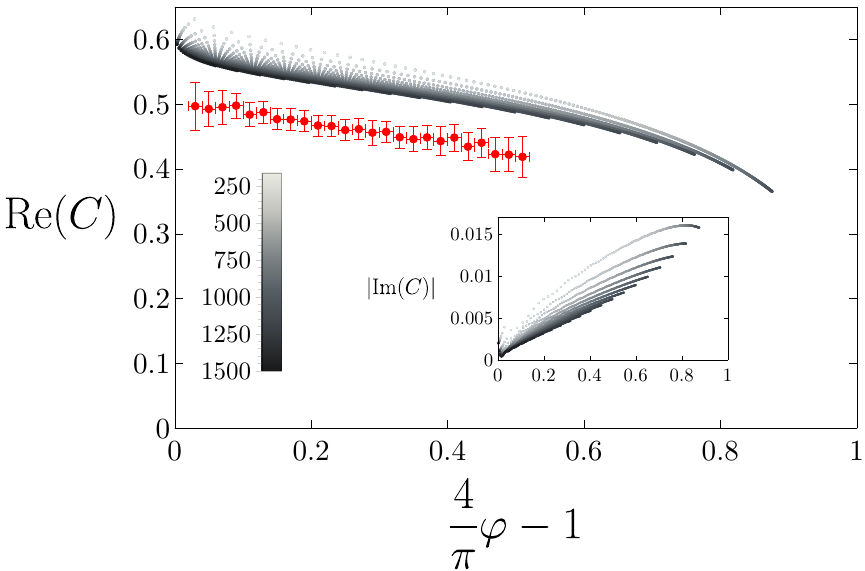}
    \caption{Dynamical connected correlation function $C\equiv \braket{\bs \sigma_j^x(t\cos\varphi)\bs \sigma^x_\ell(t\sin\varphi)}-\braket{\bs \sigma_j^x(t\cos\varphi)}\braket{\bs \sigma^x_\ell(t\sin\varphi)}$, at fixed rescaled position $z=-2$ ($z=(j-v_{\rm max}t\cos\varphi)/|t v''(\bar p)|^{\frac{1}{3}}=(\ell-v_{\rm max}t\sin\varphi)/|t v''(\bar p)|^{\frac{1}{3}}$)  with $\beta=0.75, h=0.5$. The main plot and the inset display the real part and the imaginary part, respectively. Different shades of gray correspond to different values of $t$, as indicated in the legend. The extrapolation to $t\to\infty$ {(see Section~\ref{sec:numerical} for details)} is represented by red points.}
    \label{fig:dynamical}
\end{figure}

\begin{figure}
    \centering
    \includegraphics[width=0.95\linewidth]{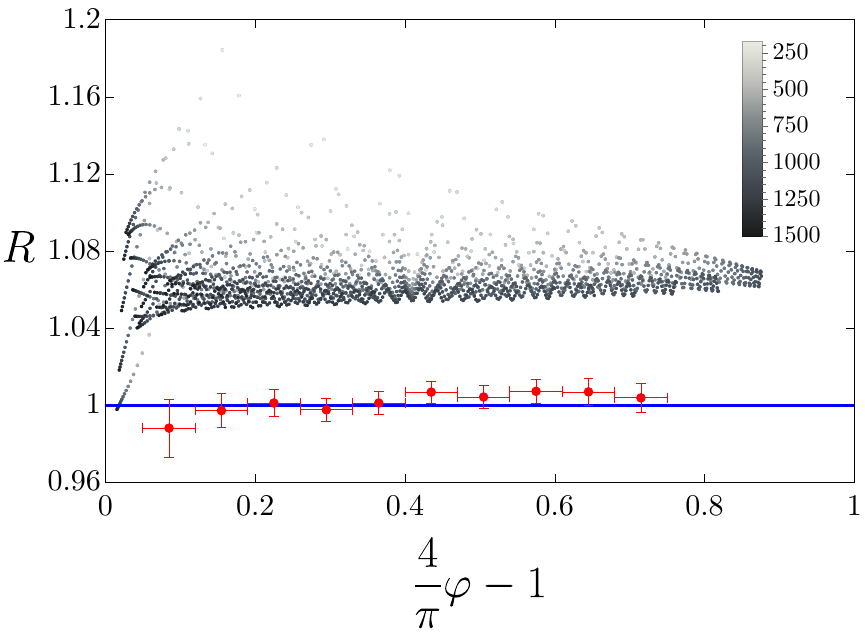}
    \caption{
    Ratio~\eqref{eq:R}, with $\bs O_j=\bs\sigma^x_j\bs\sigma^z_{j+1}$ and $(t_1,t_2)=(t\cos\varphi,t\sin\varphi)$, at fixed scaling variable $z=-2$ ($z=(j-v_{\rm max}t_1)/|t v''(\bar p)|^{\frac{1}{3}}=(\ell-v_{\rm max}t_2)/|t v''(\bar p)|^{\frac{1}{3}}$) with $\beta=0.75, h=0.5$. Different shades of gray correspond to different values of $t$, as indicated in the legend. The extrapolation to $t\to\infty$ {(see Section~\ref{sec:numerical} for details)} is represented by red points, while the solid blue line is the prediction {$R=1$. Note, for comparison, that the one-point functions in the ground state are $\braket{\sigma^x}_{\textsc{gs}}\approx 0.965$, $\braket{O}_{\textsc{gs}}\approx 0.241$.}
   }
    \label{fig:dynamicalratio}
\end{figure}

\paragraph{Non-equal time correlations.}
While our theoretical analysis  has been  focussed on equal-time correlations, we think that non-equal time correlations play a role that is too important to be completely overlooked. In equilibrium, the weak clustering of local observables across different times is an inherent part of its definition~\cite{Haag1974}. In our specific case in which relaxation occurs along a space-time curve, we find it natural to wonder what happens to non-equal time correlations in the limit of infinite time at fixed rescaled position $z$ when the times of the operators are comparable.  Here we report the results of a preliminary investigation. 
Specifically, Fig.~\ref{fig:dynamical} shows the dynamical correlation functions
\begin{equation}
    C(t\cos\varphi,t\sin\varphi)=\braket{\bs \sigma_j^x(t\cos\varphi)\bs \sigma^x_\ell(t\sin\varphi)}_{c,0},
\end{equation}
where sites {$j,\ell$ correspond to the same rescaled position, but at different times, i.e., $z_{j,t\cos\varphi}=z_{\ell,t\sin\varphi}$, with $\frac{\pi}{4}\leq \varphi<\frac{\pi}{2}$}. The extrapolation to $t\to\infty$ is obtained with the same procedure used to test stability against localised perturbations, and the meaning of the error bars is the same as in Fig.~\ref{fig:defect}. We show only the extrapolation of the real part of the correlation function, as the imaginary part is consistent with  zero  in the limit $t\to\infty$. Even if we have not worked out a theoretical prediction for the non-equal time correlations,  our extrapolations suggest quite clearly that the dynamical correlation functions are nonzero for $t\to\infty$ in the entire range of $\varphi$ investigated. This shows that the breakdown of clustering is not only in the space direction. 

The universality properties of the equal-time correlations made us conjecture that a similar result could hold true for non-equal time correlations, i.e.,  
eq.~\eqref{universal relation dynamical}. We have performed a preliminary check of this conjecture. In particular,  
 Fig.~\ref{fig:dynamicalratio} reports the ratio
{
\begin{equation}\label{eq:R}
    R=\frac{\mathrm{Re}(\braket{\bs O_j(t_1)\bs O_\ell(t_2)}_{c,0})}{\mathrm{Re}(\braket{\bs \sigma_j^x(t_1)\bs \sigma^x_\ell(t_2)}_{c,0})} \frac{\braket{\bs \sigma^x}^2_{\textsc{gs}}}{\braket{\bs O}^2_{\textsc{gs}}} \; ,
\end{equation}
}
with $\bs O_j=\bs\sigma^x_j\bs\sigma^z_{j+1}$. Even in this case the extrapolation to $t\to\infty$ agrees with the prediction {$R=1$.} 

We point out, however, that only a theoretical prediction for the late-time behaviour of non-equal time correlations could confirm or contradict these indications. Thus, we leave the behaviour of non-equal time correlations as an open question deserving further investigations.

\subsection{Interfacial region width}

Both our predictions and the numerical data that we exhibited point to the width of the interfacial region to grow as $t^{1/3}$. We discuss here the remaining dependence on the left reservoir, i.e., on $\eta$.
We have shown in Fig.~\ref{fig:mx} the profile of the magnetization for different values of the temperature of the left reservoir. The plots suggest that the interfacial region becomes wider as the temperature is decreased. This is indeed the case, as shown in the following. We define the $\epsilon$ rescaled width $\xi_\epsilon$ of the interfacial region as the length of the interval of rescaled positions $z$ for which $\mathcal{M}_\eta(z)$ is in $(\epsilon,1-\epsilon)$. 
Note that, even if parameter $\eta$ of the perturbative expansion diverges as $\beta\rightarrow 0$, the interfacial region does not shrink to zero in the limit of infinite temperature---see  Fig.~\ref{fig:magnetization_infinite_temperature}.
In the low-temperature limit, $\beta\to\infty$, on the other hand, the rescaled width diverges, consistently with the findings of Ref.~\cite{Eisler2016Universal,eisler2020Front,Delfino2022Space}. We estimate how $\xi_\epsilon$ scales with $\beta$ at fixed $\epsilon$ by truncating the perturbation theory at first order, $\mathcal{M}_\eta(z)=\exp(-\eta I_1(z))$, indeed the first-order approximation becomes exact in the limit $\beta\rightarrow\infty$ for given $z$, and it looks like it does it uniformly in $z$. Since $\eta\sim 2\exp[-2\beta \sqrt{1-h^2}]$ and $I_1(z)\sim (-z)^{3/2}$ as $z\to-\infty$  we conclude that the solution $z_1$ of $\mathcal{M}_\eta(z_1)=\varepsilon$ scales as $\exp[4\beta\sqrt{1-h^2}/3]$ as $\beta \to\infty$. On the other hand, $I_1(z)$ approaches zero faster than any power as $z\to\infty$. Thus the right boundary of the interfacial region remains close to $z=0$ and is subleading with respect to $-z_1$ as $\beta\to\infty$. From this it follows $\xi_\epsilon\sim \exp[\frac{4}{3}\beta\sqrt{1-h^2}]$. Note that the effective theory describing the edge of the lightcone does not capture the behaviour at a distance $\sim t$ from the edge, hence, by consistency, this estimation makes sense only if time is so large that $\xi_\epsilon\ll t^{2/3}$.

\section{Entanglement asymmetry}\label{s:asymmetry}

Order parameters, such as the expectation values of nonsymmetric operators, are not the only tools to investigate symmetry breaking. An alternative/complementary quantity that has been recently introduced is the ``entanglement asymmetry''~\cite{asymmetry}---cf.~Refs~\cite{Aberg2006Quantifying,White2009The}. The quantity has already received substantial attention~\cite{asym_ferro,Capizzi2023Entanglement,Ares2023Lack,Capizzi2024universal,Caceffo2024Entangled,Joshi2024,Rylands2024,Murciano2024Entanglement,Khor2024,Klobas2024Nonequilibrium,Chen2024}. Its most evident advantage is that it is only a property of the state of the (sub)system and of the symmetry group.
As remarked in Ref.~\cite{asym_ferro} in the context of global quenches in the transverse-field Ising model, the entanglement asymmetry  witnesses symmetry breaking even when the standard order parameter vanishes.

The entanglement asymmetry is defined for a density matrix $\rho_A$ and a finite symmetry group $G$ acting on it as
\begin{equation}\label{asym_def}
\Delta S_A= S(\rho_{A, G}) - S(\rho_A) ,
\end{equation}
where $S(\rho) = -\tr(\rho \log \rho)$ is the von Neumann entropy and $\rho_{A, G}$ is the state of the subsystem after applying a random element of the symmetry group with equal probabilities. That is to say, if the group has a finite dimension $|G|$, we have
\begin{equation}
    \rho_{A, G} = \frac{1}{|G|}\sum_{g \in G} g \rho_A g^ {-1}.
\end{equation}
En passant, note that this definition can be used for a generic  subsystem $A$ only if the group is generated by a local charge $Q=\sum_i Q_i$ whose density $Q_i$ has support on a single site; 
the action of an element $g\in G$ is then well defined on $\rho_A$. 
The properties of the von Neumann entropy ensure that the asymmetry is positive, bounded by $\log |G|$ and is zero if and only if $g\rho_A g^{-1} = \rho_A$ for all $g\in G$. One can conclude that  $\rho_A$ is symmetric if and only if $\Delta S_A$ is zero, whereas $\rho_A$ completely breaks the symmetry if $\Delta S_A = \log |G|$. 
Just as the von Neumann entropy is often generalised to Rényi-$\alpha$ entropies,
\begin{equation}
    S^{(\alpha)}(\rho) = \frac{1}{1-\alpha}\log(\tr(\rho^\alpha))
\end{equation}
(the von Neumann entropy corresponds to the limit $\alpha \to 1^+$), the asymmetry is naturally generalised to Rényi-$\alpha$ asymmetries $\Delta S_A^{(\alpha)}$ by replacing the von Neumann entropy with the Rényi-$\alpha$ ones in the definition \eqref{asym_def}. 

The ferromagnetic ground state of the transverse-field Ising chain breaks the $\mathbb{Z}_2$ symmetry corresponding to the group $G = \{\bs I, \bs P\}$, where $\bs I$ is the identity and $\bs P = \Pi_j \bs \sigma^z_j$ is the spin flip operator. Consequently, 
\begin{equation}\label{eq:symrdm}
\rho_{A, \mathbb Z_2} = \frac{1}{2} \rho_A + \frac{1}{2}\bs P_A \rho_A \bs P_A
\end{equation}
where $\bs P_A = \Pi_{j\in A} \sigma^z_j$, which is the even part of the reduced density matrix $\rho_A$.
\subsection{Free-fermion technique}
The $\mathbb{Z}_2$ Rényi-$\alpha$  asymmetry shares, with the order parameter correlations, the complication of requiring to compute the expectation value of odd observables, i.e., observables that anticommute with the spin flip operator ($\bs O \bs P = -\bs P \bs O)$. Consider, for example, the Rényi-$2$ asymmetry. It is convenient to decompose the reduced density matrix $\rho_A = \rho_{A, e} + \rho_{A, o}$ into an even part, $\rho_{A, e}$, and an odd part, $\rho_{A, o}$, in such a way that the former commutes  with $P_A$ and the latter anticommutes with it. Then we have
\begin{equation}
    \Delta S^{(2)}_A = \log\bigl(1 + \tfrac{\tr(\rho_{A, o}^2)}{\tr(\rho_{A, e}^2)}\bigr),
\end{equation}
where 
\begin{equation}
    \tr(\rho_{A, o}^2) = \frac{1}{2^{|A|}}\sum_{\bs O_A, \mathrm{odd}}\langle \bs{O}_A \rangle^2\, ,
\end{equation}
for a system of $|A|$ spins. Here the sum is over odd involutions, $\bs O_A^2=\bs I$. In principle, each term of the sum can be computed using cluster decomposition as we did for the order parameter. 
The number of odd observables to compute, however, grows exponentially with $|A|$, so this approach is not appropriate for large subsystems. 
Ref.~\cite{asym_ferro} proposed a trick to circumvent this issue. The idea is to express the odd part of $\rho_A$ as the product of a simple {odd }operator with a Gaussian. For example, {if $A=\llbracket l,r\rrbracket$ one can use the operator $\sigma^x_l$ acting on the first site of the subsystem, and rewrite}
\begin{equation}
\rho_{A,o} = \tr( \sigma_l^x \rho_{A})  \sigma^x_l \tilde \rho_{A}
\end{equation}
where $\tilde \rho_A$ is the normalised Gaussian
\begin{equation}
    \tilde \rho_A = \lim_{R\to \infty}\frac{\tr_B(\bs \sigma_l^x \bs \sigma_R^x  \bs \rho_e)}{\tr(\bs \sigma_l^x \bs \sigma_R^x \bs \rho_e)}\, .
\end{equation}
{Let us then use the notation $\Gamma_{A,0}$ for the correlation matrix of $\rho_A$ and $\Gamma_{A,1}$ for that of $\tilde \rho_A$. The latter can be obtained using Pfaffians (or a Schur complement, which can make the calculation faster \cite{asym_ferro})}
.
It turns out that the Rényi-$n$ entanglement asymmetry can be expressed as follows
\begin{equation}\label{eq:ent_asymm_pbc}
    \Delta S_A^{(n)} = \tfrac{1}{n-1}\log \bigl[
    \sum_{\vec \chi}c_{\vec \chi} \tfrac{\{\Sigma_{(1)}^{s_1}\Gamma_{A, \chi_1}\Sigma_{(1)}^{s_1}, \cdots, \Sigma_{(1)}^{s_n}\Gamma_{A, \chi_n} \Sigma_{(1)}^{s_n}\}}{\{\Gamma_{A, 0}, \overset{n}{\dots}, \Gamma_{A, 0}\}}\bigr],
\end{equation}
where {$\Sigma_{(1)}$ is a diagonal involution with the first diagonal element equal to $-1$ and the rest of them equal to $1$}, the sum is over $\vec \chi = (\chi_1, \cdots, \chi_n)$ with $\chi_j\in \{0, 1\}$, $s_i = \sum_{j=1}^i \chi_j$ and $c_{\vec \chi} = \frac{1+(-1)^{s_n}}{2}\langle \sigma^x_l \rangle^{s_n}$.
In \eqref{eq:ent_asymm_pbc} we are using the bracket notations of Ref.~\cite{Fagotti2010disjoint}, according to which the trace of the product of $n$ Gaussian density matrices $\rho_1, \cdots, \rho_n$ with  correlation matrices $\Gamma_1, \cdots, \Gamma_n$ reads 
$
    \tr(\rho_1 \cdots \rho_n) = \{\Gamma_1,\cdots ,\Gamma_n\}
$. Such products can be defined  recursively as follows~\cite{Fagotti2010disjoint}
\begin{equation}
    \{\Gamma_1, \cdots, \Gamma_n\} = \{\Gamma_1, \Gamma_2\} \{\Gamma_1 \times \Gamma_2, \Gamma_3, \cdots, \Gamma_n\},
\end{equation}
where $\Gamma_1 \times \Gamma_2$ stands for the correlation matrix of the normalised product of two Gaussians with correlation matrices $\Gamma_1$ and $\Gamma_2$ and is given by
\begin{equation}\label{eq:corProd}
\Gamma_1\times\Gamma_2=\mathrm I-(\mathrm I-\Gamma_2)\tfrac{\mathrm I}{\mathrm I+\Gamma_1\Gamma_2}(\mathrm I-\Gamma_1)\, ;
\end{equation}
$\{\Gamma_1, \Gamma_2\}$ is the product of the eigenvalues of $\frac{\mathrm{I} + \Gamma_1 \Gamma_2}{2}$ with halved degeneracy (each eigenvalue is doubly degenerate because the correlation matrices are skew symmetric):
\begin{equation}
    \{\Gamma_1, \Gamma_2\} = \prod_{\mu \in \spec(\Gamma_1 \Gamma_2)/2} \tfrac{1+\mu}{2}.
\end{equation}
Representation~\eqref{eq:ent_asymm_pbc} is appropriate to compute the R\'enyi-$n$ asymmetry for small enough integer $n>1$. We are not aware of an analogous free-fermion representation of the von Neumann asymmetry. 

\subsection{Conjecture and numerical checks}
\begin{figure}
    \centering
    \includegraphics[width=0.95\linewidth]{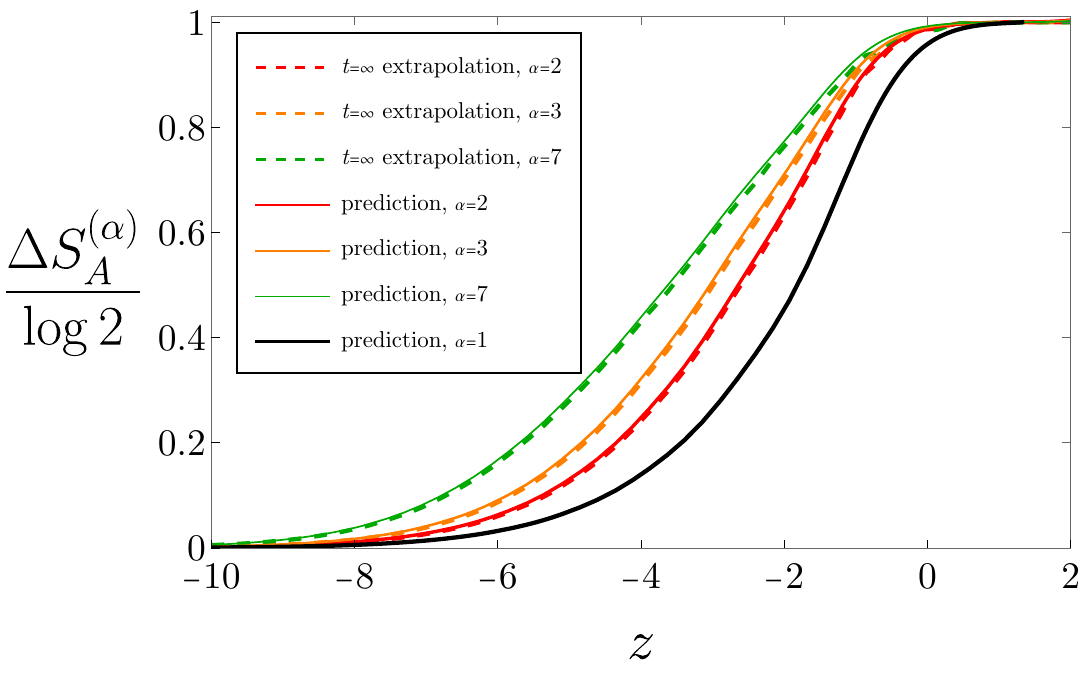}
    \caption{Rényi-$\alpha$ and von Neumann ($\alpha=1$) asymmetries of a comoving subsystem $A$ whose left and right boundaries have rescaled coordinates $-12$ and $z$. The {dashed} lines correspond to the $t\to \infty$ limit obtained by extrapolation of the numerical data, while the {solid} lines correspond to prediction \eqref{asym_formula}.}
    \label{fig:asymmetry}
\end{figure}
For systems $A_t(z-\epsilon,z)$ with  $\epsilon \ll 1$, it is straightforward to derive an analytical formula for the Rényi-$\alpha$ and von Neumann asymmetry using that the expectation value of every local even observable approaches the ground state value, while the expectation value of every odd observable is multiplied by the scaling function $\mathcal M_\eta(z)$. This implies that the reduced density matrix of  $A_t(z-\epsilon,z)$ is
\begin{equation}
    \rho_{A} \underset{\epsilon \ll 1}{=} \frac{1+\mathcal M_\eta(z)}{2}\rho^{\GS}_{A} + \frac{1-\mathcal M_\eta(z)}{2}\bs P_A \rho^{\GS}_{A}\bs P_A\; ,
\end{equation}
where $\rho^{\GS}_{A}$ is the corresponding reduced density matrix in the ground state and $A$ stands for $A_t(z-\epsilon,z)$. 
Since there is no string order associated with the aforementioned spin-flip symmetry in the ferromagnetic ground state, for large enough subsystems (such that $\log 2 - \Delta S_A\ll 1$),  $\|\rho^{\GS}_{A}\bs P_A \rho^{\GS}_{A}\|$ is exponentially smaller than $\|\rho^{\GS}_{A} \rho^{\GS}_{A}\|$, hence we get
 \begin{equation}\label{asym_formula}
    \Delta S_A^{(\alpha)} = \log 2 - H_\alpha(\mathcal M_\eta(z))+o(|\epsilon t^{\frac{1}{3}}|^{-\gamma})
\end{equation}
as $1\ll\epsilon t^{\frac{1}{3}}\ll t^{\frac{1}{3}}$, for any $\gamma>0$,
where $H_\alpha(x)$ is the Rényi-$\alpha$ entropy function of a single bit of information
\begin{equation}
    H_\alpha(x) = \frac{1}{1-\alpha}\log\left(\left(\frac{1+x}{2}\right)^\alpha+\left(\frac{1-x}{2}\right)^\alpha\right)\, .
\end{equation}
This is expected to remain true also in the limit $\alpha\rightarrow 1^+$, in which $H_\alpha$  approaches the Shannon entropy function $H_1(x)=-\frac{1+x}{2}\log\frac{1+x}{2}-\frac{1-x}{2}\log\frac{1-x}{2}$.

It is reasonable to expect that, subleading corrections apart, the asymptotic result~\eqref{asym_formula} captures also the late time behaviour of a subsystem $A_t(z_l, z)$ even when $z-z_l$ doesn't approach zero. An argument supporting this conclusion follows. The starting point is to assume that \eqref{asym_formula} is correct in the limit $z_l\rightarrow z$. Let us then identify the effect of moving the left boundary $z_l$ to the left. First we observe that $\bs P_A\rho_A^{\GS}\bs P_A$ corresponds to the contribution coming from having an odd number of excitations to the right hand side of $A$: The string operator $\bs P$, indeed, maps one symmetry breaking ground state into the other, which is what a quasilocalised (semilocal) excitation does to the left hand side of its position. This is quantitatively captured by the scaling function at the right boundary of the subsystem, independently of how large the subsystem is. Thus, we argue that $\rho_A$ keeps the same structure as in \eqref{asym_formula}, provided to replace $\rho_A^{\GS}$ with the appropriate density matrix. Since we still do not expect string order in the state, we end up with \eqref{asym_formula} even when $z-z_l$ does not approach $0$.

We have checked this conjecture considering semi-infinite subsystems $A_t(-\infty,z)$. Fig.~\ref{fig:asymmetry} shows a comparison between conjecture~\eqref{asym_formula} and numerical data. The extrapolation to $t\to\infty$ is obtained assuming $\sim t^{-1/3}$ and $\sim t^{-2/3}$ corrections and perfectly agrees with the prediction (the tiny but visible discrepancy can be traced back to the truncation of the perturbative expansion used to approximate the scaling function). 

We remark that the strong link between the asymptotic behaviour captured by \eqref{asym_formula} and the asymptotics of the order parameter~\eqref{eq:sx} is generally absent after global quenches, where, instead, the $\mathbb{Z}_2$ entanglement asymmetry of a subsystem can be order $1$ while the local longitudinal magnetization approaches zero~\cite{asym_ferro}.

\section{Wigner-Yanase skew information}\label{s:skew}
\begin{figure}[t]
    \centering
  \includegraphics[width=1\linewidth]{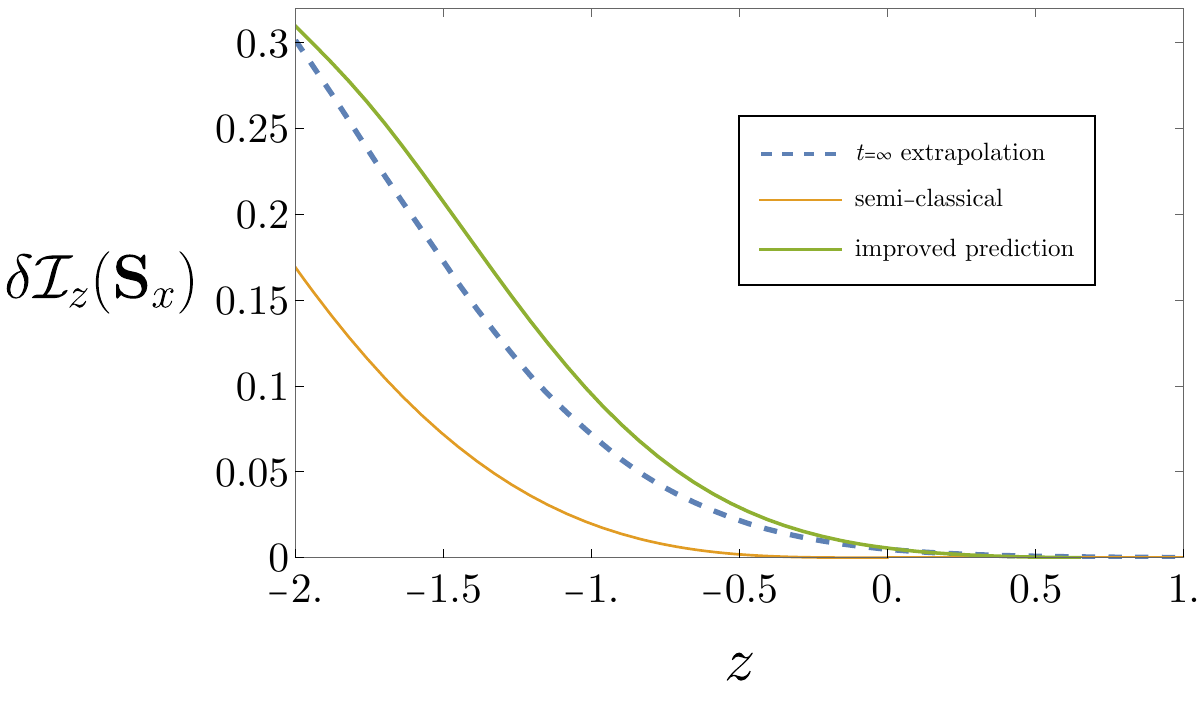}
    \caption{Skew information profile  for the semi-infinite subsystem with $\beta=0$.}
    \label{fig:skew_info}
\end{figure}
The protocol that we studied is special since, generally, the size of fully correlated clusters remains finite even at late times when time evolving under translationally invariant Hamiltonians with local interactions.  
In the interfacial region, on the other hand, we have shown that such size diverges as $t^{1/3}$. We are aware of other one-dimensional settings in which similar exceptional properties are triggered by localized or semi-localized perturbations:
\begin{itemize}
\item[-] in symmetry breaking ground states~\cite{Zauner2015,Eisler2016Universal,Eisler2018Hydrodynamical,Ferro2025Kicking}.
\item[-] in symmetric states in the presence of semilocal conservation laws~\cite{Fagotti2022Global}.
\item[-] in jammed states of kinetically constrained systems~\cite{Bidzhiev2022Macroscopic,Zadnik2022Measurement,Fagotti2024Quantum}.
\item[-] in separable---or, with both low bipartite and low multipartite entanglement---quantum scars~\cite{Bocini2023growing}.
\end{itemize}
Refs~\cite{Bocini2023growing,Fagotti2024Quantum, Ferro2025Kicking}, in particular, point out that the corresponding clusters are not only full-range classically correlated, but also full-range quantum correlated. 

We find it reasonable to identify the quantum part of the variance of an operator $\bs O$ with a quarter of the quantum Fisher information $F_{\rho}(\bs O)$, indeed Refs~\cite{Toth2013Extremal,Yu2013Quantum} showed that $\frac{1}{4}F_{\rho}(\bs O)$ is the convex roof of the variance
\begin{equation}
\frac{1}{4}F_{\rho}(\bs O)=\inf_{\{p,\ket{\Psi}\}}\sum_i p_i[\braket{\Psi_i|\bs O^2|\Psi_i}-\braket{\Psi_i|\bs O|\Psi_i}^2]
\end{equation}
where the infimum is taken over all possible convex decompositions of the density matrix, $\rho$, in pure states, $\rho=\sum_i p_i\ket{\Psi_i}\bra{\Psi_i}$. We remark that Ref.~\cite{Frerot2016Quantum} proposed an alternative definition 
\begin{equation}
\Delta^2_Q\bs O=-\frac{1}{2}\int_0^1d\alpha\tr\left([\bs O,\rho^\alpha][\bs O,\rho^{1-\alpha}]\right)\, ,
\end{equation}
but there are constants $\kappa_{1,2}$ such that $\kappa_1\Delta^2_Q\bs O\leq \frac{1}{4}F_{\rho}(\bs O)\leq \kappa_2\Delta^2_Q\bs O$, hence the two definitions are equivalent. Since we are not aware of an efficient way to compute the quantum Fisher information in a subsystem of an almost Gaussian state, such as the one under investigation, we have opted for computing an equivalent---in the same sense as before---quantity: the Wigner-Yanase skew information
\begin{equation}
I_\rho(\bs O)=-\frac{1}{2}\tr\left([\bs O,\rho^{1/2}]^2\right)\, .
\end{equation}
Since we already showed that  the two-point function of $\bs \sigma^x$ does not cluster in the interfacial region, we consider $I_{\rho_A}(\bs S^x[A])$ in subsystems $A$ in the right edge comoving frame (here and in the following $A$ is a shorthand for $A_t(z,z')$ with generic $z$ and $z'$).
This quantity can be computed in a Gaussian state with free fermion techniques, as follows. The first step is to express the skew information as
\begin{equation}
I_{\rho_A}(\bs S^x_A)=\frac{1}{4}\sum_{\ell,n\in A}\tr\left((\rho_A-\kappa^{(\ell)}_A \tilde\rho_A^{(\ell)})\bs \sigma_\ell^x\bs \sigma_n^x \right)
\end{equation}
where $\kappa^{(\ell)}_A$ equals $\tr(\rho_A^{1/2}\bs \sigma_\ell^x\rho_A^{1/2}\bs \sigma_\ell^x)$ and $\tilde\rho_A^{(\ell)}$ is the normalised Gaussian 
\begin{equation}
\tilde\rho_A^{(\ell)}=(\kappa^{(\ell)}_A)^{-1}\rho_A^{1/2}\bs \sigma_\ell^x\rho_A^{1/2}\bs \sigma_\ell^x\, .
\end{equation}
Both $\kappa^{(\ell)}_A$ and the correlation matrix of $\tilde\rho_A^{(\ell)}$ can be easily expressed in terms of the correlation matrix $\Gamma_A$ of $\rho_A$ within the formalism developed in Ref.~\cite{Fagotti2010disjoint}. Specifically, if $\Sigma_{(2\ell-1)}$ is the diagonal involution with the first $2\ell-1$ diagonal elements equal to $-1$ and the rest of them equal to $1$, we have
\begin{equation}
\kappa^{(\ell)}_A=\sqrt{\det\left|\left(\mathrm I+\Bigl(\tfrac{1-\sqrt{1-\Gamma_A^2}}{\Gamma_A} \Sigma_{(2\ell-1)}\Bigr)^2\right)/2\right|}
\end{equation}
and
\begin{equation}
\tilde \Gamma_A^{(\ell)}=\tfrac{1-\sqrt{1-\Gamma_A^2}}{\Gamma_A}\times \left(\Sigma_{(2\ell-1)}\tfrac{1-\sqrt{1-\Gamma_A^2}}{\Gamma_A}\Sigma_{(2\ell-1)}\right)\, ,
\end{equation}
where $\Gamma_1\times\Gamma_2$ was defined in \eqref{eq:corProd}.
We are then left with a sum of  expectation values of $\bs\sigma_\ell^x\bs\sigma_{n}^x$ in Gaussian states, as in eq.~\eqref{two point correlator pfaffian}.

To highlight the breakdown of clustering in the interfacial region, we study the increase of the Wigner-Yanase skew information $\Delta I_z[\bs S^x]$ in  half-infinite subsystems $A\equiv A_t(z,\infty)$
\begin{equation}
\Delta I_z(\bs S^x)=I_{\rho_{A}(t)}(\bs S^x[A])-I_{\rho_{A}(0)}(\bs S^x[A])\, .
\end{equation}
This is finite by virtue of the Lieb-Robinson bounds as the contribution from the operators sufficiently far to the right from the edge of the lightcone cancels with the corresponding one at the initial time. 
{We point out that, as explained in Section \ref{s:asymmetry}, $\rho_A$ is not Gaussian, indeed the Gaussian state associated to $\rho_A$ is the symmetrized state of Eq.~\eqref{eq:symrdm}. Because of its convexity properties, the Wigner-Yanase skew information in that state is only a lower bound to the skew information of the symmetry-broken state (the skew information of $\rho_A(t)$ is the same as that of $\bs P_A \rho_{A} (t)\bs P_A$). However, considering the semi-infinite subsystem up to $+\infty$ allows us to ignore this issue. Indeed, the eigenstates of $\rho_{A}(t)$ and $\bs P_A \rho_{A} (t)\bs P_A$ become exactly orthogonal since they correspond to two different symmetry-breaking ground states on a semi-infinite region, i.e. extending sufficiently towards the right from the edge of the light-cone. The skew information of the Gaussian state is thus the sum of the two identical contributions coming from $\rho_{A}(t)$ and $\bs P_A \rho_{A} (t)\bs P_A$, divided by $2$.}
{In practice,} $A_t(z,\infty)$ could be replaced by $A_t(z,z_{\mathrm{r}})$ with $z_{\mathrm{r}}>0$ up to exponentially small corrections in $z_{\mathrm{r}}$. The effective size of the latter subsystem is $\sim t^{1/3}$. In view of this, we compute $\delta\mathcal I_z(\bs S^x)\equiv |v''(\bar p)t|^{-2/3}\Delta I_z(\bs S^x)$, which can approach a nonzero value in the limit $t\rightarrow\infty$ only if quantum correlations do not cluster within the interfacial region. An example is reported in Fig.~\ref{fig:skew_info}, in which we show the $z$ dependence of an extrapolation to $t\rightarrow\infty$. 
It is evident that $\delta \mathcal I_z(\bs S^x)$ remains nonzero for values of $z$ in the interfacial region, proving that the region is full-range quantum correlated.

\section{Physical interpretation}\label{sec:physical_int}
We propose here a basic semiclassical description that provides a good approximation for the order parameter correlations and a fair one for the Wigner-Yanase skew information. 
Since around the edge we are close to zero temperature, as a starting point we take the semiclassical theory developped by Sachdev and Young in Ref.~\cite{Sachdev1997Low}.
The excitations are then described by classical particles traveling at constant speed. 
Odd operators are semilocal with respect to the elementary excitations, that is to say, their expectation values change sign when a particle crosses their position.  
This results in the following semiclassical representation for the one-point function of a local odd operator $\bs O(z)$ at rescaled position $z$ 
\begin{equation}\label{eq:one-point}
\braket{\bs O(z)}\overset{\mathrm{s.c.}}{=}\braket{\bs O(z)}_{\GS}\sum_{j}(-1)^jP(\nenearrows_{\!\! j}^{(z,\infty)})\, ,
\end{equation}
where $P(\nenearrows_{\!\! j}^{(z_1,z_2)})$ stands for the probability that there are $j$ particles in the (mobile) subsystem in the right edge comoving frame within rescaled positions $z_1$ and $z_2$. (Here and in the following the probabilities are defined for a fixed large time $t$.) 
Analogously, the two-point function is represented as
\begin{multline}\label{eq:two-points}
\braket{\bs O(z_1)\bs O(z_2)}\overset{\mathrm{s.c.}}{=}\\
\braket{\bs O(z_1)\bs O(z_2)}_{\GS}\sum_{j}(-1)^jP(\nenearrows_{\!\! j}^{(z_1,z_2)})\, .
\end{multline}
We now compute such probabilities treating the excitations as classical particles. 
Since only the fastest particles can reach the interfacial region, it is convenient to rewrite the trajectory of the particles in rescaled coordinates: $z=z_0-\frac{1}{2}q^2$, where we introduced a space variable, $z_0=\frac{x_0}{|v''(\bar p) t |^{1/3}}$, which becomes continuous in the limit of infinite time, and $x_0$ labels the chain site where the particle was originated.
Since  the initial state has a finite correlation length, we can assume that, at large enough time, particles originated at different $z_0$ are completely uncorrelated. Note that a fixed tiny interval $\Delta z_0$ can be the origin of a number of particles that scales as $t^{1/3}$. Such particles are uncorrelated as long as their distance is larger enough than the local correlation length in the initial state. 
We shall make the approximation that there are no correlations between the classical particles at the initial time, which can be rephrased as that only one particle can be originated in the infinitesimal interval $(z_0,z_0+dz_0)$. The probability to have one excitation with a finite rescaled momentum $q$ at $z_0$ at the initial time is asymptotically independent of $q$ and given by 
\begin{equation}
P(\leftindex_{(z_0,z_0+dz_0)}{\nearrow}_{\!\!(q,q+d q)})=\tfrac{1-e^{-\eta}}{4\pi}\theta_H(-z_0)
dz_0 dq
\end{equation}
A particle originated at $z_0$ has crossed $z$ before time $t$ only if its rescaled momentum $q$ is smaller than $\sqrt{2(z_0-z)}$ in absolute value, therefore 
the joint probability that a particle was produced at $z_0$ and,  at time $t$, is to the right hand side of rescaled position $z$ reads
\begin{equation}
P(\leftindex_{(z_0,z_0+dz_0)}{\nearrow}^{(z,\infty)})=\tfrac{1-e^{-\eta}}{\sqrt{2}\pi}\sqrt{z_0-z}\theta_H(z_0-z)dz_0\, .
\end{equation}
Since particles are assumed to be uncorrelated, this directly gives the probability to have $0$ particles to the right hand side of rescaled position $z$
\begin{multline}
P(\nenearrows_{\!\! 0}^{(z,\infty)})=e^{-\int^0_{-\infty} P(\leftindex_{(z_0,z_0+dz_0)}{\nearrow}^{(z,\infty)})}\\
=e^{-\frac{\sqrt{2}(1-e^{-\eta})\, .}{3\pi}(-z)^{\frac{3}{2}}\theta_H(-z)}\, .
\end{multline}
Note that for $z\geq 0$  the probability is estimated to become exactly $1$ because the maximal velocity is a hard upper bound for classical particles. 
Analogously we have
\begin{multline}\label{eq:Pj}
P(\nenearrows_{\!\! j}^{(z,\infty)})\\
=\tfrac{\left[\int^0_{-\infty}P(\leftindex_{(z_0,z_0+dz_0)}{\nearrow}^{(z,\infty)})\right]^j}{j!}e^{-\int^0_{-\infty}P(\leftindex_{(z_0,z_0+dz_0)}{\nearrow}^{(z,\infty)})}\\
=\tfrac{[\frac{\sqrt{2}(1-e^{-\eta})}{3\pi}(-z)^{\frac{3}{2}}]^j}{j!}e^{-\frac{\sqrt{2}(1-e^{-\eta})}{3\pi}(-z)^{\frac{3}{2}}}\, .
\end{multline}
The semiclassical approximation of the one-point function, and, in turn, of the one-point scaling function is then readily obtained---cf. \eqref{eq:one-point} and \eqref{eq:Pj}
\begin{equation}\label{scaling_sc}
\mathcal M_\eta(z)\overset{\mathrm{s.c.}}{=} P(\nenearrows_{\!\! 0}^{(z,\infty)})^2=e^{-\frac{2\sqrt{2}(1-e^{-\eta})}{3\pi}(-z)^{\frac{3}{2}}}\, .
\end{equation}
By repeating the same steps for the $2$-point function---eq.~\eqref{eq:two-points}---we find 
\begin{multline}
\mathcal M_\eta(z_1,z_2)\overset{\mathrm{s.c.}}{=} \frac{P(\nenearrows_{\!\! 0}^{(z_1,\infty)})^2}{P(\nenearrows_{\!\! 0}^{(z_2,\infty)})^2}\\
=e^{-\frac{2\sqrt{2}(1-e^{-\eta})}{3\pi}[(-z_1)^{\frac{3}{2}}-(-z_2)^{\frac{3}{2}}]}\, .
\end{multline}
We are also in the position to estimate the variance per unit $|v''(\bar p)t|^{2/3}$ of  an extensive odd operator $\bs O=\sum_\ell \bs O_\ell$ in  $A_t(z_1,z_2)$:
\begin{multline}
\lim_{t\rightarrow\infty}\tfrac{1}{|v''(\bar p)t|^{2/3}\braket{\bs O}_{\GS}^2}\mathrm{Var}_t(\bs O[A_t(z_1,z_2)])\\
\overset{\textrm{s.c.}}{\approx}4\int_{z_1}^{z_2}d z_1'e^{-\frac{2\sqrt{2}(1-e^{-\eta})}{3\pi}(-z_1')^{\frac{3}{2}}}\\
\int_{z_1'}^{z_2}d z_2'\sinh\Bigl(\tfrac{2\sqrt{2}(1-e^{-\eta})}{3\pi}(-z_2')^{\frac{3}{2}}\Bigr)\\
\sim \tfrac{16\sqrt{2}(1-e^{-\eta})}{105\pi}[ 2 (-z_1)^{\frac{7}{2}} + 7 z_1 (-z_2)^{\frac{5}{2}} + 5 (-z_2)^{\frac{7}{2}}]
\end{multline}
for $z_1<z_2\leq 0$; the result for $z_2$ larger than $0$ is obtained by replacing $z_2$ with $0$.
Notwithstanding the simplicity of the derivation, these semiclassical approximations are rather good for small $\eta$ and they also capture the leading asymptotic behaviour of $I_1(z)$ for large $(-z)$.

While the physical interpretation that we provided is essentially classical, we can use it also to estimate the Wigner-Yanase skew information, which captures purely quantum correlations. At first sight, this might sound impossible; the secret to get a nonzero result is to complement the semiclassical description by additional information, which is beyond the classical theory. 
Specifically, for a half-infinite subsystem in the edge comoving frame with the left boundary at rescaled position $z$ close to the edge of the lightcone, we found the following assumptions to be consistent with the numerical observations.
\begin{itemize}
\item[-] The contributions coming from more than one particle in the subsystem are negligible.
\item[-] The single particle sector associated with a particle originated at a given $z_0$ is quantum coherent.
\item[-] Contributions from particles originated at different $z_0$ are incoherent, and that is the only source of incoherence.  
\end{itemize}
Under these assumptions, the Wigner-Yanase skew information equals a quarter of the quantum Fisher information and can be expressed in terms of the classical probabilities as follows
\begin{multline}
\frac{I_{\rho_{A_t(z,\infty)}}(\bs O[A_t(z,\infty)])}{|v''(\bar p) t|^{2/3}\braket{\bs O}_{\GS}^2}\\
\overset{\mathrm{s.c.}}{=}8\int_{-\infty}^0 P(\nenearrows_{\!\! 1}^{(z,\infty)},\leftindex_{(z_0,z_0+dz_0)}{\nearrow}^{(z,\infty)})\\
\int_z^\infty dz_1\int_{z_1}^\infty dz_2\tfrac{P(\leftindex_{(z_0,z_0+dz_0)}{\nearrow}^{(z_2,\infty)})}{P(\leftindex_{(z_0,z_0+dz_0)}{\nearrow}^{(z,\infty)})}\\
\Bigl(1-\tfrac{P(\leftindex_{(z_0,z_0+dz_0)}{\nearrow}^{(z_1,\infty)})}{P(\leftindex_{(z_0,z_0+dz_0)}{\nearrow}^{(z,\infty)})}\Bigr)\, .
\end{multline}
Here $P(\nenearrows_{\!\! 1}^{(z,\infty)},\leftindex_{(z_0,z_0+dz_0)}{\nearrow}^{(z,\infty)})$ stands for the probability that there is only one particle in the subsystem and that that particle was originated in $(z_0,z_0+dz)$. The rest of the expression comes from expressing the variance in terms of probabilities as done above under the condition that the unique particle that is present in the subsystem was originated at $z_0$ (with the level of approximation we are working on, we can make the identification $P(\leftindex_{(z_0,z_0+dz_0)}{\nearrow}^{(z_1,z_2)})\sim P(\leftindex_{(z_0,z_0+dz_0)}{\nearrow}^{(z_1,\infty)})-P(\leftindex_{(z_0,z_0+dz_0)}{\nearrow}^{(z_2,\infty)})$). 

Within the semiclassical picture we have
\begin{multline}
P(\nenearrows_{\!\! 1}^{(z,\infty)},\leftindex_{(z_0,z_0+dz_0)}{\nearrow}^{(z_1,z_2)})\\
=\tfrac{1-e^{-\eta}}{\sqrt{2}\pi}[(z_0-z_1)^{\frac{1}{2}}-(z_0-z_2)^{\frac{1}{2}}]e^{-\frac{\sqrt{2}(1-e^{-\eta})}{3\pi}(- z)^{\frac{3}{2}}} d z_0\, .
\end{multline}
Thus, the probability that the particle originated at $z_0$ is in $(z_1,z_2)$ given that there is a single particle to the right hand side of $z$ is
\begin{equation}
\tfrac{P(\nenearrows_{\!\! 1}^{(z,\infty)},\leftindex_{(z_0,z_0+dz_0)}{\nearrow}^{(z_1,z_2)})}{P(\nenearrows_{\!\! 1}^{(z,\infty)})}=\left[(\tfrac{z_0-z_1}{z_0-\bar z})^{\frac{1}{2}}-(\tfrac{z_0-z_2}{z_0-\bar z})^{\frac{1}{2}}\right] dz_0
\end{equation}
We can then express the contribution to the one- and two-point function of $\bs O$  restricted  to particles originated at $z_0$ and knowing that there's only one particle in the subsystem as follows
\begin{align}
\tfrac{\braket{\bs O(z)}^{(z_0)}}{\braket{\bs O}_{\GS}}&=1-2(\tfrac{z_0-z }{z_0- z })^{\frac{1}{2}}\\
\tfrac{\braket{\bs O(z_1)\bs O(z_2)}^{(z_0)}}{\braket{\bs O}_{\GS}^2}&=1-2(\tfrac{z_0-z_1}{z_0- z})^{\frac{1}{2}}+2(\tfrac{z_0-z_2}{z_0- z})^{\frac{1}{2}}
\end{align}
The corresponding variance per unit $|v''(\bar p)t|^{2/3}$ equals $\frac{16}{45}(z_0- z)^2\braket{\bs O}_{\GS}^2$, and hence the Wigner-Yanase skew information is estimated as follows
\begin{equation}
\tfrac{I_{\rho_A(t)}[\bs O[A]]}{|v''(\bar p) t|^{2/3}\braket{\bs O}_{\GS}^2}\overset{\mathrm{s.c.}}{=}\tfrac{32}{315}\tfrac{1-e^{-\eta}}{\sqrt{2}\pi}(-z)^{\frac{7}{2}}e^{-\frac{\sqrt{2}(1-e^{-\eta})}{3\pi}(-z)^{\frac{3}{2}}}
\end{equation}
where $A$ is a shorthand for $A_t(z,\infty)$.
The comparison with numerical data for not too large $-z$ is qualitatively fair but quantitatively poor---cf. Fig.~\ref{fig:skew_info}. One of the reasons behind it is that the semiclassical approximation for the probabilities is already leaving out quantum contributions that are of the same order as the contributions considered here. To some extent, we can reintroduce them by hand.
Specifically, we propose to identify the probability to have zero particles in a subsystem (in the right edge comoving frame) $P(\nenearrows_{\!\! 0}^{(z_1,z_2)})$ with the overlap between the reduced density matrix and the corresponding one in the initial state. The calculation follows the same steps as for the order parameter correlation, and indeed the result is very similar:
\begin{equation}
P(\nenearrows_{\!\! 0}^{(z_1,z_2)})=\mathcal M_{\tilde \eta}(z_1,z_2)\, ,\ P(\nenearrows_{\!\! 0}^{(z,\infty)})=\mathcal M_{\tilde \eta}(z_1)\; ,
\end{equation}
with $\tilde\eta=\log\frac{2}{1+e^{-\eta}}$. We keep the same assumptions as in the semicalssical approximation;  for example, we have
\begin{multline}
    P(\leftindex_{(z_0,z_0+dz_0)}{\nearrow}^{(z,\infty)}) = -\partial_{z_0}\log \left(P(\nenearrows_{\!\! 0}^{(z-z_0,\infty)})\right) dz_0\\
    = - \partial_{z_0}\log \left(\mathcal{M}_{\tilde \eta}(z-z_0)\right) dz_0\; .
\end{multline}
The result at the leading order of the perturbation theory according to which $\mathcal{M}_{\tilde \eta}(z)\sim e^{-\tilde \eta I_1(z)}$  is  the following refined approximation
\begin{multline}
\frac{I_{\rho_{A_t(z,\infty)}(t)}[\bs O[A_t(z,\infty)]]}{|v''(\bar p) t|^{2/3}\braket{\bs O}_{\GS}^2}\sim \\
4\tilde\eta e^{-\tilde\eta I_1(z)}\int^{\infty}_{ z} d z_0 \left[\frac{[ I_1(z_0)]^2}{I_1'(z_0)}+2 \int_{z_0}^\infty d z'  I_1(z')\right]\, .
\end{multline}
which is the formula that we compared against numerical data in Ref.~\cite{Maric2024Macroscopic}. The reader can appreciate how much this refined prediction outperforms the semiclassical approximation by revisiting Fig.~\ref{fig:skew_info}. As a final note, we disclose that, within the interval of rescaled positions shown in the figure, the residual discrepancy between prediction and extrapolation arises almost entirely from the error coming from approximating the probability of having zero particles to the right of the rescaled position $z$ by $e^{-\tilde\eta I_1(z)}$.

\section{Conclusions}

In this work we have derived analytic expressions for the time evolution of one- and two-point functions in the transverse field Ising chain after joining a symmetry breaking ground state with a disordered state. We identified an interfacial region near the right edge of the light cone, scaling as $t^{1/3}$, where correlations converge to universal functions. Within this region, we demonstrated that correlations are full-range, and we computed the Wigner-Yanase skew information of the order parameter to confirm that these strong correlations include a quantum component. We have studied the entanglement asymmetry of subsystems within the interfacial region and obtained an analytic prediction that exhibits the same degree of universality as the order parameter correlation functions. We have proposed a semiclassical theory based on the  Sachdev-Young one, which helps one understand  the behaviour of correlations, asymmetry, and even quantum Fisher information in the interfacial region.

In light of the exceptional classical and quantum properties near the edge of the lightcone that we pointed out, we think that it is compelling to supplement our investigation with  the study of other quantities,
such as the entanglement negativity~\cite{Vidal2002,Plenio2005,Coser2014negativity,Wen2015,Eisler2015On,Gruber2020,Eisler2023,Bertini2022EntanglementNegativity},  the mutual and the tripartite information~\cite{Eisler2014,Kormos2017Temperature,Alba2018Entanglement,Alba2019QuantumInformationScrambling,Parez2022,Caceffo2023} (we are particularly curious about the residual value of the latter~\cite{Maric2022Universality,Maric2023universality,Maric2023universality2,Maric2024Entanglement}), the Markov gap~\cite{Zou2021,Berthiere2024}, and the entanglement Hamiltonian~\cite{Dalmonte2022,DiGiulio2019,Rottoli2024}.

In Ref.~\cite{Maric2024Macroscopic}  we argued that the phenomenology that we unveiled extends to interacting integrable systems, but a detailed analysis is still lacking, and it is unclear what to expect in the presence of integrability breaking interactions. To shed light on interacting integrable systems, it might be of interest to obtain some of the results of this work within a form factor approach~\cite{Konik2008,Granet2020}. 
And in view of the universality of the behaviours, we think that it could be instructive to derive the results again within the framework of a quantum field theory~\cite{Delfino2022Space}.
In more generic systems, numerical methods based on tensor networks could help recognize the crucial properties of the interfacial region. In that respect, we wonder whether some aspects  could also be addressed  through quantum circuits. While we have considered time evolution in an isolated system, the protocol that we studied has a direct analogue in open quantum systems~\cite{Fazio2024}, where one could imagine to prepare a semi-infinite system in a symmetry breaking ground state and put it in contact with a disordered reservoir at the boundary described by a Lindblad master equation.  Finally, in light of recent advances in quantum quenches~\cite{Gibbins2024,Yamashika2024} and interface dynamics~\cite{Balducci2022,Balducci2023,Pavesic2024constrained} in higher dimensions, 
we wonder whether the problem of joining an ordered state with a disordered reservoir could be effectively addressed also
in two and three dimensions.

\begin{acknowledgments}
We thank Alberto Rosso for useful discussions. 
This work was supported by the European Research Council under the Starting Grant No. 805252 LoCoMacro.
\end{acknowledgments}

\bibliography{references}

\end{document}